\documentclass[12pt,preprint,nofootinbib,noshowkeys,noshowpacs,longbibliography,
superscriptaddress,tightenlines,floatfix,aps,prc]{revtex4-2}
\usepackage{amsmath,amsfonts,amsthm,amssymb}
\usepackage{graphics,graphicx}
\usepackage{color}
\usepackage{esint}
\usepackage{siunitx}
\DeclareSIUnit{\angstrom}{\textup{\AA}}
\usepackage{dsfont}	
\definecolor{darkblue}{RGB}{0,0,196}
\definecolor{darkgreen}{RGB}{0,120,0}

\newcommand{\vect}[1]{\boldsymbol{\mathbf{#1}}}

\newcommand*\dif{\mathop{}\!\mathrm{d}}

\usepackage{mathrsfs}              
\usepackage{mathtools}

\def\beq{\begin{equation}}
\def\eeq{\end{equation}}
\def\st{\begin{equation}}
\def\stp{\end{equation}}
\def\ba{\begin{eqnarray}}
\def\ea{\end{eqnarray}}

\usepackage[colorlinks=true,linktocpage=true,linkcolor=darkblue,citecolor=red,urlcolor=darkblue]{hyperref}

\begin{document}
\preprint{}
 
    \title{A note on the canonical approach to hydrodynamics and linear response theory}
    \author{Luca Martinoia}
    \email{luca.martinoia@ge.infn.it}
    \affiliation{Dipartimento di Fisica, Università di Genova, via Dodecaneso 33, I-16146, Genova, Italy}
    \affiliation{I.N.F.N. - Sezione di Genova, via Dodecaneso 33, I-16146, Genova, Italy}

    \author{Rajeev Singh}
    \email{rajeev.singh@e-uvt.ro}
    \affiliation{Center for Nuclear Theory, Department of Physics and Astronomy, Stony Brook University, Stony Brook, New York, 11794-3800, USA}
    \affiliation{School of Physical Sciences, National Institute of Science Education and Research, An OCC of Homi Bhabha National Institute, Jatni-752050, India}
    \affiliation{Department of Physics, West University of Timisoara, Bd.~Vasile P\^arvan 4, Timisoara 300223, Romania}
	\date{\today}
	\bigskip
\begin{abstract}
This note provides a comprehensive examination of the various approaches to formulating relativistic hydrodynamics, with a particular emphasis on the canonical approach. Relativistic hydrodynamics plays a crucial role in understanding the behavior of fluids in high-energy astrophysical phenomena and heavy-ion collisions. The canonical approach is explored in detail, highlighting its foundational principles, mathematical formulations, and practical implications in modeling relativistic fluid dynamics. Following this, we delve into the linear response theory, elucidating its relevance in the context of hydrodynamics. We analyze the response of relativistic fluids to external perturbations, discussing the theoretical framework and key results. This dual focus aims to bridge the gap between theoretical foundations and practical applications, offering a robust perspective on the dynamic interplay between relativistic hydrodynamics and linear response theory.
\end{abstract}
     
\date{\today}
\maketitle
\newpage
\tableofcontents

\section{Regime of hydrodynamics and its applications}
\label{sec:hydrodynamic_regime}
Hydrodynamics, although well-explored \cite{Landau:FluidMechanicsVolume}, remains a forefront area of research due to its persistent numerical and theoretical challenges. Its universality renders it invaluable for describing diverse phenomena across disciplines such as astrophysics \cite{Shibata:GeneralRelativisticViscous,Faber:HydrodynamicsNeutronStar,Balbus:InstabilityTurbulenceEnhanced}, condensed and soft matter physics \cite{Lucas:HydrodynamicsElectronsGraphene,Narozhny:ElectronicHydrodynamicsGraphene,Hartnoll:TheoryNernstEffect,Baggioli:ColloquiumHydrodynamicsHolography,Lucas:HydrodynamicTheoryThermoelectric,Amoretti:2019buu,Toner:FlocksHerdsSchools,Armas:2024iuy,Amoretti:2024obt}, and high-energy physics, exemplified by studies on quark-gluon plasma at RHIC and LHC \cite{Teaney:2009qa,Palni:2024wdy,Arslandok:HotQCDWhite,Becattini:2024uha,Heinz:CollectiveFlowViscosity,Rocha:2023ilf,Sorensen:2023zkk}.

Recent years have seen a surge of interest in hydrodynamics, catalyzed by the rediscovery of fluid/gravity duality in holography \cite{Rangamani:GravityHydrodynamicsLectures,Hartnoll:HolographicQuantumMatter}. This revival has spurred investigations into various facets of theoretical hydrodynamics, including the classification of dissipative superfluid terms \cite{Bhattacharya:TheoryFirstOrder}, advancements in second- \cite{Baier:RelativisticViscousHydrodynamics,Bhattacharyya:NonlinearFluidDynamics,Romatschke:RelativisticViscousFluid} and third-order hydrodynamics \cite{El:ThirdorderRelativisticDissipative,Grozdanov:ConstructingHigherorderHydrodynamics}, parity-odd fluids \cite{Jensen:ParityViolatingHydrodynamicsDimensions,Lier:PassiveOddViscoelasticity,Lucas:PhenomenologyNonrelativisticParityviolating}, quantum anomalies \cite{Son:HydrodynamicsTriangleAnomalies,Jensen:TriangleAnomaliesThermodynamics,Amoretti:2022vxq}, spin hydrodynamics \cite{Gallegos:HydrodynamicsSpinCurrents,Becattini:SpinTensorIts,Florkowski:2018fap,Florkowski:2019qdp,Florkowski:2021wvk,Singh:2020rht,Singh:2022uyy,Speranza:2020ilk,Bhadury:2020puc,Bhadury:2021oat,Bhadury:2022ulr}, viscoelastic and viscoplastic fluids with broken translation symmetry (both spontaneous and explicit) \cite{Amoretti:2017axe,Amoretti:2017frz,Amoretti:2018tzw,Armas:ApproximateSymmetriesPseudoGoldstones,Armas:HydrodynamicsPlasticDeformations,Armas:HydrodynamicsChargeDensity,Baggioli:ColloquiumHydrodynamicsHolography,Amoretti:2021lll,Amoretti:2021fch,Amoretti:2022acb}, fluctuating hydrodynamics developed from an action principle \cite{Kovtun:EffectiveActionRelativistic,Grozdanov:ViscosityDissipativeHydrodynamics,Glorioso:LecturesNonequilibriumEffective,Haehl:FluidManifestoEmergent,Haehl:EffectiveActionRelativistic}, comprehensive classification of hydrodynamic transport properties \cite{Haehl:AdiabaticHydrodynamicsEightfold,Haehl:EightfoldWayDissipation}, quasihydrodynamics \cite{Baggioli:QuasihydrodynamicsSchwingerKeldyshEffective,Grozdanov:HolographyHydrodynamicsWeakly}, Hydro+ extension incorporating slow modes near critical points \cite{Stephanov:HydroHydrodynamicsParametric}, fracton hydrodynamics \cite{Glodkowski:HydrodynamicsDipoleconservingFluids,Grosvenor:HydrodynamicsIdealFracton,Guo:FractonHydrodynamicsTimereversal}, hydrodynamics with higher-form symmetries \cite{Das:HigherformSymmetriesAnomalous,Grozdanov:GeneralizedGlobalSymmetries,Armas:ApproximateHigherformSymmetries}, steady-state hydrodynamics~\cite{Amoretti:2022ovc,Amoretti:2024jig,Brattan:2024dfv}, and hydrodynamics via variational principle~\cite{Jackiw:2004nm,Nair:2011mk,Karabali:2014vla,Monteiro:2014wsa,Nair:2020kjg}.

Before delving deeper, it's crucial to define hydrodynamics as a universal low-energy effective field theory applicable to many-body thermal systems. It captures the collective macroscopic dynamics of conserved charges or other low-energy modes in the regime of long-wavelengths and small-frequency $\omega, \vect{k} \ll T$, with $T$ denoting the system's temperature. To understand how hydrodynamics emerges, we follow the arguments presented in \cite{Glorioso:LecturesNonequilibriumEffective}.

In a quantum many-body system, at zero temperature, the low-energy theory is characterized by massless quasiparticle excitations above the ground state, which typically exhibit long-lived behavior and give rise to phenomena such as Fermi-liquid behavior. However, in a system at thermal equilibrium, there exists a substantial reservoir of gapless quasiparticles. Any new excitation introduced into this system quickly loses coherence within the bath over a timescale $\tau$ and length scale $l$ determined by microscopic dynamics. This rapid loss of coherence effectively suppresses the quasiparticles, leading to their rapid decay and the system's prompt thermalization.

As a consequence, on macroscopic scales, the system can be considered to be in local thermodynamic equilibrium. This concept is best grasped from a coarse-grained perspective, where the system is divided into small volume elements. Each element is sufficiently small to be treated as point-like within the description of the effective theory, yet large enough at the microscopic level to uphold a well-defined thermodynamic limit.
\begin{figure}[!htp]
	\centering
	\includegraphics[width=0.4\textwidth]{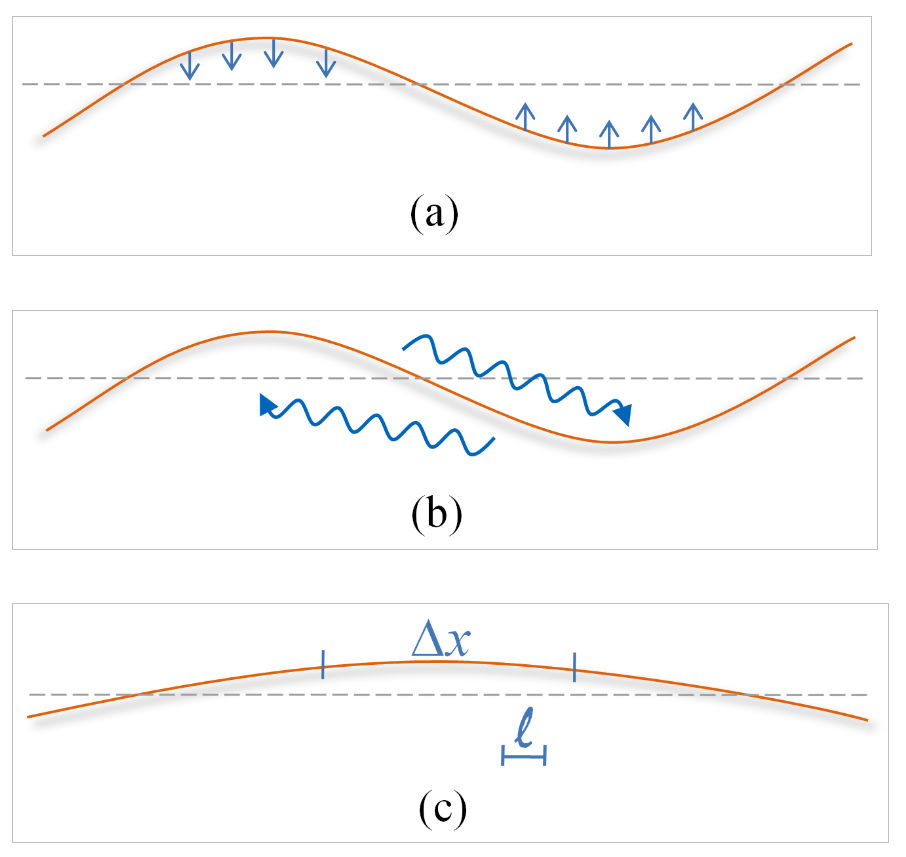}
	\caption{(a) Microscopic degrees of freedom rapidly thermalize to equilibrium. (b) Conserved charges, however, cannot be (locally) destroyed and (slowly) diffuse. (c) The characteristic microscopic length scale $l$ must be significantly smaller than the length over which conserved charges vary~\cite{Glorioso:LecturesNonequilibriumEffective}.}
	\label{fig:hydrodynmic_regime}
\end{figure}

In this regime, the dynamics at low energies are dictated by conserved charges, which cannot be locally eliminated but can only diffuse from regions with an excess of charge to those with a deficit. This diffusion process occurs at a rate significantly slower than local microscopic thermalization (see Figure~\ref{fig:hydrodynmic_regime}). Consequently, the system reaches the hydrodynamic limit, where a low-energy theory emerges to describe long-wavelength and long-timescale dynamics governed by the local conservation of charges. This framework may also encompass other low-energy collective degrees of freedom, as observed in superfluid phases.

The distinction in scales between microscopic thermalization and macroscopic diffusion, crucial for the emergence of hydrodynamics, is particularly pronounced in phases that are strongly coupled and devoid of impurities. In these systems, the scattering time among quasiparticles becomes exceedingly small~\cite{Martinoia:2024cbw}.

We hope that this note would be beneficial for early graduate students and interested researchers.
\section{Approaches to formulate hydrodynamics}
\subsection{Canonical approach}\label{sec:canonical_approach}
In the canonical approach, (relativistic) hydrodynamics is interpreted as a perturbative expansion in the Knudsen number, $\text{Kn} = l/L \ll 1$, where $L$ denotes the system size or the macroscopic length scale over which thermodynamic quantities change, and $l$ represents the microscopic length scale. 
Precisely, we can define the derivative operator $l\partial \sim \text{Kn}$, which serves as the parameter for the perturbative expansion. From this perspective, hydrodynamics is founded on the following constraints:
\begin{enumerate}
    \item A derivative expansion,
    \item The continuous, broken, and discrete symmetries,
    \item Second law of thermodynamics in its local form.
\end{enumerate}
This approach treats hydrodynamics as a formal power series, where the zeroth order $(\partial^0)$ corresponds to a non-dissipative perfect fluid and higher-order ($\partial^1$ or above) corrections introduce dissipation by adding entropy to the system. Ideally, as with any effective theory, including higher-order terms should improve the accuracy of the description within a certain range of validity. However, significant evidence suggests that the hydrodynamic series does not converge \cite{Heller:HydrodynamicsGradientExpansion,Heller:2021oxl}. Despite this, it can be Borel resummed to provide an all-order hydrodynamic description.

Based on the preceding discussion, it is evident that hydrodynamics ceases to be a valid effective theory at length scales shorter than the microscopic mean free path $l$. Each term in the derivative expansion, starting from the first order, is associated with a transport coefficient that characterizes the fluid dynamics' non-universal aspects. These coefficients are functions of thermodynamic quantities and cannot be directly determined if one considers hydrodynamics alone; their values also rely on the underlying microscopic (kinetic) theory. They can be calculated using Kubo formulae through various methods such as lattice calculations, kinetic theory, holography, and quantum field theory.

Despite hydrodynamics' success in describing the physics of many different systems, as a classical theory, it faces foundational problems.
One significant issue is that the standard hydrodynamic equations are acausal and unstable, since they include solutions in which the group velocity diverges and becomes superluminal~\cite{Hiscock:GenericInstabilitiesFirstorder,Hiscock:LinearPlaneWaves,Speranza:ChallengesSolvingChiral,Heller:2022ejw,Heller:2021yjh}. This prevents numerical evolution of the relativistic hydrodynamic equations from a set of initial conditions on a time-slice, compromising the predictable nature of the theory.

The solution to this problem was first proposed by Maxwell and Cattaneo for diffusion \cite{Maxwell:IVDynamicalTheory,Cattaneo:SullaConduzioneCalore}, and later extended by M\"uller, Israel, and Stewart (MIS) \cite{Mueller:ParadoxonWaermeleitungstheorie,Israel:TransientRelativisticThermodynamics,Israel:NonstationaryIrreversibleThermodynamics}. Their approach involves introducing soft ultraviolet (UV) cutoffs, denoted as $\tau_i$, to regulate the high-momentum behavior of the hydrodynamic fluctuations. Specifically, defining $\pi^{\mu\nu}$ as the dissipative shear tensor, $\Pi$ as the dissipative bulk pressure, and $\nu^\mu$ as the charge flux, the MIS theory recommends altering the constitutive relations in the dynamical equations as follows
\begin{subequations}
\begin{align}
\pi^{\mu\nu} &= -\eta \left( \nabla^\mu u^\nu + \nabla^\nu u^\mu - \frac{2}{3} \Delta^{\mu\nu} \nabla_\alpha u^\alpha \right) - \tau_\pi u^\alpha \partial_\alpha \pi^{\mu\nu}\,, \\
\Pi &= -\zeta \nabla_\alpha u^\alpha - \tau_\Pi u^\alpha \partial_\alpha \Pi\,, \\
\nu^\mu &= -\kappa \nabla^\mu T - \tau_\nu 
u^\alpha \partial_\alpha \nu^\mu \,,
\end{align}
\end{subequations}
where $\eta$ is the shear viscosity, $\zeta$ is the bulk viscosity, $\kappa$ is the thermal conductivity, $u^\mu$ is the fluid velocity, $\Delta^{\mu\nu}$ is the projector orthogonal to $u^\mu$, and $T$ is the temperature. The parameters $\tau_\pi$, $\tau_\Pi$, and $\tau_\nu$ are relaxation times associated with the respective transport coefficients.

The corrections introduced by this method only come into play at very high frequencies. However, the hydrodynamic equations are now strongly hyperbolic, allowing for numerical solutions given a set of initial conditions. They also feature a well-behaved thermal equilibrium state that is both stable and causal. The new parameters $\tau_i$ are linked to UV degrees of freedom. In MIS theory, a new non-hydrodynamic mode $\omega(\vect{k}=0)\sim-{i}/{\tau_\pi}$ is found, which decays quickly \cite{Romatschke:RelativisticFluidDynamics}.

In recent years, a new proposal has emerged to address the problem of causality and stability in relativistic hydrodynamics \cite{Bemfica:NonlinearCausalityGeneral,Bemfica:CausalityExistenceSolutions,Kovtun:FirstorderRelativisticHydrodynamics,Hoult:StableCausalRelativistic,Abboud:CausalStableFirstorder,Biswas:2022hiv}. This approach takes advantage of an ambiguity inherent in any hydrodynamic description. Fluid variables are not well-defined out of equilibrium, leading to ambiguity up to derivative corrections, known as frame redefinitions~\cite{Bhattacharyya:2023srn,Bhattacharyya:2024tfj,Bhattacharyya:2024ohn}. Using this method, it is possible to identify a frame (referred to as the BDNK frame) in which first-order relativistic hydrodynamics becomes strongly hyperbolic leading to causal and stable equations. However, this comes at the cost of having more complex constitutive relations.

MIS solution requires us to work with second-order hydrodynamics to achieve full consistency \cite{Baier:RelativisticViscousHydrodynamics}. However, this approach introduces new fundamental issues. 
This is because of the presence of hydrodynamic modes $\omega(\vect{k})$, which are long-lived solutions to the conservation equations that can lead to instabilities in the system by violating the assumption of local thermal equilibrium~\cite{Kovtun:LecturesHydrodynamicFluctuations}.

Classical hydrodynamics traditionally focuses solely on the dissipative part of the system, neglecting the effects of thermal fluctuations. However, fluctuations should always be considered alongside dissipative effects, as dictated by the fluctuation-dissipation theorem. This can be addressed by including fluctuations as noise in the equations, thereby transforming hydrodynamics into a proper EFT, following the Martin-Siggia-Rose formalism \cite{Martin:StatisticalDynamicsClassical,Landau:HydrodynamicFluctuations,Hohenberg:TheoryDynamicCritical,Kovtun:EffectiveActionRelativistic,Kovtun:LecturesHydrodynamicFluctuations}.

A top-down approach to dissipative fluctuating hydrodynamics has been developed \cite{Glorioso:LecturesNonequilibriumEffective,Grozdanov:ViscosityDissipativeHydrodynamics,Harder:ThermalFluctuationsGenerating,Haehl:TwoRoadsHydrodynamic}, utilizing the closed time-path formalism of Schwinger and Keldysh \cite{Chou:EquilibriumNonequilibriumFormalisms}. This method operates at the full non-linear level and allows access to $n$-point correlators.

Beyond linear order, modes can interact with themselves through thermal fluctuations, leading to the renormalization of transport coefficients compared to their bare, linear-response values. This renormalization depends on the UV scale $\Lambda_\text{UV}$, beyond which hydrodynamics is no longer valid, and results from interactions between modes near the cutoff. These interactions, which can even occur in equilibrium because of thermal fluctuations, may lead to the breakdown of the hydrodynamic derivative expansion \cite{Kovtun:LecturesHydrodynamicFluctuations}. 

From a technical standpoint, this occurs because interactions between modes alter the retarded correlators by non-analytical terms, that scale like $\mathcal{O}(\omega^{3/2})$ in $3+1$ dimensions~\cite{DeSchepper:NonexistenceLinearDiffusion,Kovtun:StickinessSoundAbsolute}, however in $2+1$ dimensions it is a logarithmic correction \cite{Forster:LargedistanceLongtimeProperties}. 
Hence, these corrections are effectively more significant than those in first-order hydrodynamics, which are linear in $\mathcal{O}(\omega)$, yet smaller than second-order corrections, which scale as $\mathcal{O}(\omega^2)$, hence, these corrections cannot be captured by local derivative terms.
This non-analytic behavior is related to long-time tail effect: the non-analytic $\omega^{3/2}$ dependence of, for example, $3+1$ dimensional hydrodynamics corresponds to a real-time evolution as $t^{3/2}$ of the response function \cite{Pomeau:TimeDependentCorrelation}, which decays slowly compared to exponential decay.

Furthermore, as previously mentioned, hydrodynamics ceases to be a sensible effective field theory at length scales shorter than the microscopic scale $l$. However, this assumption, central to the canonical approach, is challenged by observations that hydrodynamics seems to work well as a model for systems where the separation of scales is not sharply defined. This has spurred a significant amount of research aiming to understand this ``unreasonable effectiveness of hydrodynamics.'' Investigations in this area have concentrated on resummation methods for all-order hydrodynamics \cite{Heller:HydrodynamicsGradientExpansion,Romatschke:RelativisticFluidDynamics,Santos:DivergenceChapmanEnskogExpansion,Heller:HydrodynamizationKineticTheory} and the hydrodynamic attractor \cite{Romatschke:FluidDynamicsFar,Bu:LinearizedFluidGravity}, as well as an action-like description of hydrodynamics via the Schwinger-Keldysh formalism \cite{Glorioso:LecturesNonequilibriumEffective,Haehl:TwoRoadsHydrodynamic,Grozdanov:ViscosityDissipativeHydrodynamics}.

We emphasize that the problems associated with causality, stability, and fluctuations arise when interpreting the equations of hydrodynamics as a set of classical partial differential equations (PDEs) to solve for the real-time evolution of the fluid. These issues stem from high-frequency, high-wavevector modes that are far from the hydrodynamic regime (which is not expected to hold at microscopic length scales). 
After providing a better context on the state of hydrodynamics, we will focus on the linear response theory of fluids in the hydrodynamic regime, concentrating on the $\omega,\vect{k}\rightarrow0$ properties of small fluctuations around a global thermodynamic equilibrium state, for which the above mentioned problems are not relevant.
\subsection{Kinetic theory}
One of the earliest methods stems from kinetic theory \cite{Huang:StatisticalMechanics,Denicol:MicroscopicFoundationsRelativistic}. At first glance, this might seem strange since hydrodynamics is often associated with strongly-coupled phases of matter that thermalize quickly, whereas kinetic theory is typically seen as describing weakly-coupled theories. However, this apparent contradiction dissipates when we realize that hydrodynamics is highly universal: in the presence of a single species of quasiparticles (as is also often the case in kinetic theory) and no impurities, there is nothing to inhibit the onset of hydrodynamics. It naturally emerges at long wavelengths even at small couplings.

MIS hydrodynamics arises naturally from kinetic theory \cite{Israel:TransientRelativisticThermodynamics,Denicol:MicroscopicFoundationsRelativistic,Denicol:DerivationTransientRelativistic}, providing a justification for the timescales $\tau_i$ introduced phenomenologically.
Recently, there have been developments in kinetic theory that extend its application to hydrodynamics, encompassing various hydrodynamic frames \cite{Rocha:NovelRelaxationTime,Hoult:CausalFirstorderHydrodynamics,Basar:2024qxd}. This microscopic approach also offers direct formulations of transport coefficients using microscopic and thermodynamic parameters.
\subsection{Schwinger-Keldysh formalism}
Hydrodynamics is a potent tool, but from a microscopic perspective, the arguments used before to explain its emergence are purely heuristic and not entirely satisfactory. 
The canonical formulation lacks systematic inclusion of thermal and quantum fluctuations, which must be separately incorporated in bottom-up approaches. These fluctuations are particularly crucial in contexts such as turbulence, chaotic systems, and the renormalization of transport coefficients (long-time tails).

Another aspect pertains to the theory's validity regime. In classical hydrodynamics, this is constrained by the scale separation condition $l \ll L$. However, experiments in heavy-ion collisions indicate that this limit is overly stringent. Recent formulations of fluid dynamics based on action principles and all-order hydrodynamics resummation techniques have been able to transcend this paradigm.

Finally, the symmetries and thermodynamic constraints are imposed phenomenologically in classical hydrodynamics. In contrast, in the effective-action approach, these constraints arise directly from fundamental properties of the dissipative effective action, providing the theory with a stronger theoretical foundation.

From a theoretical standpoint, there has been a longstanding interest in developing field theory action-like methods, based on concepts like path integrals, to describe dissipative systems such as hydrodynamics \cite{Lazo:ActionPrincipleActiondependent,Dubovsky:EffectiveFieldTheory,Dubovsky:EffectiveFieldTheorya}. While ideal fluids and some non-dissipative aspects of hydrodynamics are already been incorporated by action-like principles \cite{Jensen:HydrodynamicsEntropyCurrent,Banerjee:ConstraintsFluidDynamics}, fully incorporating dissipative corrections from path integral methods remains an active area of research \cite{Jain:SchwingerKeldyshEffectiveField,Armas:EffectiveFieldTheory,Jain:EffectiveFieldTheory,Baggioli:QuasihydrodynamicsSchwingerKeldyshEffective}.

In the past years, the Schwinger-Keldysh formalism has been successfully used to develop hydrodynamics and thermal dissipative systems, from an action-like principle \cite{Haehl:EffectiveActionRelativistic,Glorioso:LecturesNonequilibriumEffective,Grozdanov:ViscosityDissipativeHydrodynamics}. This formulation is based on the realization that describing thermal dissipative systems requires recasting the standard problem of statistical field theory on a closed time path, leading to a doubling of the hydrodynamic degrees of freedom (advanced and retarded fields depending on which branch of the time contour they inhabit). After integrating out the UV degrees of freedom, one arrives at a Wilsonian effective action that must satisfy proper constraints, such as unitarity and $\mathbb{Z}_2$ dynamical KMS symmetry. This formulation yields an effective action encompassing all details of hydrodynamics' constitutive relations and conservation equations. Unlike the canonical method, it inherently integrates thermal (and potentially quantum) fluctuations at the complete nonlinear level, aligning with the fluctuation-dissipation theorem.
\subsection{Fluid-gravity duality}
This approach is rooted in the gauge-gravity duality, which finds its basis in string theory \cite{Maldacena:LargeLimitSuperconformal,Gubser:GaugeTheoryCorrelators,Witten:SitterSpaceHolography}. It takes various forms, but the simplest and most practical one, not relying on quantum gravity and strings, is a duality between a strongly-coupled (large 't Hooft coupling) large-$N$ QFT in $d$ dimensions on the boundary, and a theory of classical gravity in $d+1$ dimensions within bulk Anti de Sitter space. Specifically, a black hole within the interior of AdS space gives rise to thermodynamics in the dual boundary theory, with its temperature and entropy determined by the black hole's Gibbons-Hawking temperature and Bekenstein–Hawking entropy.

Prior to the discovery of holography, upon realizing that black holes are thermal systems, it was quickly inferred that the analogy extends beyond thermal equilibrium to generic hydrodynamic perturbations of the event horizon, indicating that black holes are dissipative systems, encapsulated in the membrane paradigm \cite{Thorne:BlackHolesMembrane}. This correspondence becomes clearer in AdS space \cite{Iqbal:UniversalityHydrodynamicLimit,Policastro:AdSCFTCorrespondence,Policastro:AdSCFTCorrespondencea}, forming the basis for the fluid-gravity duality.

Holography has demonstrated that the membrane paradigm analogy extends beyond linear perturbations, allowing for the fluid-gravity correspondence to operate at the full nonlinear level \cite{Bhattacharyya:NonlinearFluidDynamics}. This enables the derivation of not only the full set of hydrodynamic correlators but also the hydrodynamic equations \cite{Hartnoll:HolographicQuantumMatter,Ramallo:IntroductionAdSCFT,Rangamani:GravityHydrodynamicsLectures,Natsuume:AdSCFTDuality}. 

Holography is used, alongside hydrodynamics, to study strongly-coupled systems from two different angles and to cross-check the accuracy of hydrodynamic effective theories. Historically, many intriguing features of hydrodynamics (such as the effect of anomalies on fluids) were initially uncovered using holographic methods before being confirmed in hydrodynamics. Furthermore, akin to kinetic theory, holographic models provide the non-universal aspects of fluid description, such as the equation of state and transport coefficients, which can be used to establish bounds on transport coefficients, exemplified by the KSS bound $\eta/s\geq{1}/{4\pi}$ \cite{Kovtun:HolographyHydrodynamicsDiffusion}.
\section{Canonical approach to hydrodynamics}
\label{chapter:hydrodynamics_linear_response}
In this section, we'll concentrate on the relativistic normal fluid without any superfluid component in $d+1$ spacetime dimensions. However, the approach we'll discuss applies to fluids with any symmetry content, such as Galilean, Carrollian, boost-agnostic, and so on. Our discussion will primarily adhere to~\cite{Kovtun:LecturesHydrodynamicFluctuations}.
\subsection{Constitutive relations}
As we've discussed, hydrodynamics deals with the long-wavelength dynamics of conserved charges. According to Noether's theorem, conservation laws arise from the continuous symmetries of the underlying theory, which implies the existence of conserved currents. In our case, we'll focus on a charged relativistic fluid, which means we assume not only Poincaré symmetry (boosts, rotations, and translations) but also an internal $\mathrm{U(1)}$ symmetry, leading to a conserved vector current (e.g., baryon number, electric charge, etc.).

Symmetry under spacetime translations gives rise to a conserved symmetric stress-energy tensor (or in other words, energy-momentum tensor) $T^{\mu\nu}$, which couples to the external spacetime metric $g_{\mu\nu}$ and can be described as the derivative of the action with respect to the metric
\begin{equation}
	T^{\mu\nu}=\frac{2}{\sqrt{-g}}\frac{\delta S}{\delta g_{\mu\nu}}\,.
\end{equation}
The energy-momentum tensor derived from Noether's (first) theorem may not always be symmetric, however, using above definition the stress-energy tensor is automatically symmetric~\footnote{However, in a recent study~\cite{Singh:2024qvg} it is argued that the symmetric energy-momentum tensor can be obtained directly from Noether's second theorem instead of using Noether's first theorem.}. 
However, following Belinfante's improvement procedure, it's possible to include extra antisymmetric terms to $T^{\mu\nu}$ that are conserved and lead to a symmetric stress-energy tensor.
In this work, we will deal with only symmetric $T^{\mu\nu}$. It's important to note that if spin-angular momentum is a conserved quantity independent of the total angular momentum conservation, one may have to consider antisymmetric corrections to the stress-energy tensor~\cite{Gallegos:HydrodynamicsSpinCurrents,Singh:2022uyy}.

The conservation law for stress-energy tensor is
\begin{equation}
\partial_\mu T^{\mu\nu}=0\,.
\label{eqn:stress_energy_conservation}
\end{equation}
Computing the conserved currents related to the other spacetime symmetries such as boosts and rotations gives $\mathcal{M}^{\mu\nu\alpha}=x^\mu T^{\nu\alpha}-x^\nu T^{\mu\alpha}$. This current is automatically conserved due to \eqref{eqn:stress_energy_conservation} and the relativistic Ward Identity $T^{\mu\nu}=T^{\nu\mu}$.

Moving on to internal symmetries, for a global $\mathrm{U(1)}$ symmetry, this implies the existence of a conserved current $J^\mu$
\begin{equation}
\partial_\mu J^\mu=0 \,.
\label{eqn:current_conservation}
\end{equation}
We remark that it's indeed plausible to include higher-form symmetries leading to the conservation of totally antisymmetric currents with more than one index~\cite{Das:HigherformSymmetriesAnomalous,Armas:ApproximateHigherformSymmetries}.

It is crucial to emphasize that \eqref{eqn:current_conservation} and \eqref{eqn:stress_energy_conservation} are significant equations of motion for the conserved charges, which act as the dynamical fields in hydrodynamics. Although these equations are identically satisfied on-shell for a specific microscopic action $S$, in the context of hydrodynamics, they encapsulate the theory's dynamical essence.

Without further assumptions, it's not possible to solve the conservation equations. By simply counting the degrees of freedom, a symmetric stress-energy tensor has $(d+1)(d+2)/2$ components and the $\mathrm{U(1)}$ current has $d+1$. However, we only have $d+1$ equations from \eqref{eqn:stress_energy_conservation} and one equation from \eqref{eqn:current_conservation}, which is not enough to uniquely determine the system's evolution.

Hydrodynamics simplifies the problem by assuming that $T^{\mu\nu}$ and $J^\mu$ can be expressed in terms of $d+2$ local fields, known as hydrodynamic fields or variables. In relativistic fluid dynamics, these independent fields typically include: a local temperature $T(x)$, a local chemical potential $\mu(x)$, and a local fluid four-velocity $u^\mu(x)$ satisfying $u \cdot u = -1$.

In the grand canonical ensemble, the equilibrium of a thermal system is described by a density operator $\hat{\rho}$, which is proportional to the exponential of the conserved charges \cite{Israel:ThermodynamicsRelativisticSystems}. In our case, these conserved charges are described by the four-momentum operator $\hat{P}^\mu$ and the particle number operator $\hat{N}$. These operators are given by integrating the stress-energy tensor $\hat{T}^{\mu\nu}$ and the current $\hat{J}^\mu$ over some spacelike hypersurface $\Sigma_\mu$ respectively. 
The density operator can be expressed as~\cite{Haehl:AdiabaticHydrodynamicsEightfold}
\begin{equation}\label{eqn:thermal_density_operator}
	\hat\rho=\frac{1}{\text{Tr}\ e^{\beta_\mu \hat P^\mu+\Lambda_\beta \hat N}}e^{\beta_\mu \hat P^\mu+\Lambda_\beta \hat N}\,,
\end{equation} 
where $\beta^\mu$ (a time-like four-vector) is the the ratio of fluid velocity over temperature and $\Lambda_\beta$ (a scalar) is the ratio of chemical potential over temperature. These are two Lagrange multipliers.
The exponential term involving the conserved charges arises from the fact that in equilibrium, the expectation values of these charges are conserved, and the density operator describes the statistical distribution of these charges.

In hydrodynamics, we depart from the uniform equilibrium state characterized by \eqref{eqn:thermal_density_operator}, and we consider temperature, chemical potential, and fluid velocity varies slowly with spacetime. The key idea is that while the operators $\hat{T}^{\mu\nu}$ and $\hat{J}^\mu$ are always well-defined (for a given theory), their averages $T^{\mu\nu}=\langle\hat{T}^{\mu\nu}\rangle$ and $J^\mu=\langle\hat{J}^\mu\rangle$ can be written in terms of equilibrium quantities when these are slowly varying functions of spacetime.

It's worth noting that although a specific state of equilibrium breaks boost symmetry by selecting a particular timelike vector $\beta^\mu$, there are various equivalent states of equilibrium with different spatial velocities. Therefore, it is the state that breaks the symmetry, not the theory itself, which remains fully Lorentz covariant. Indeed, it can be explicitly demonstrated that the Goldstone boson corresponding to the spontaneous breaking of boost symmetry is not an additional dynamical field; it must be identified with the fluid velocity \cite{Armas:CarrollianFluidsSpontaneous}. This demonstrates that hydrodynamics respects the underlying Lorentz symmetry of the theory.

In discussing the constitutive relations, we aim to express the stress-energy tensor and the $\mathrm{U(1)}$ current in terms of the hydrodynamic fields in a way that respects the symmetries of the theory. In relativistic hydrodynamics, it's convenient to decompose these tensors with respect to the timelike vector $u^\mu$, as this enables us to work with quantities that are clearly identifiable as Lorentz covariant objects. We consider flat Minkowski spacetime with metric $\eta_{\mu\nu}=\text{diag}(-1,1,\dots,1)$ and define the symmetric projector orthogonal to the four-velocity as $\Delta^{\mu\nu}=\eta^{\mu\nu}+u^\mu u^\nu$. The decomposition reads~\cite{Eckart:ThermodynamicsIrreversibleProcesses,Kovtun:LecturesHydrodynamicFluctuations}
\begin{subequations}\label{eqn:tensor_current_standard_decomposition}
\begin{align}
	T^{\mu\nu}&=\mathcal{E}u^\mu u^\nu+\mathcal{P}\Delta^{\mu\nu}+\mathcal{Q}^{\mu} u^{\nu}+\mathcal{Q}^{\nu} u^{\mu}+\mathcal{T}^{\mu\nu}\,,\\
	J^\mu&=\mathcal{N}u^\mu+\mathcal{J}^\mu\,,
\end{align}
\end{subequations}
where $\mathcal{E}$, $\mathcal{P}$ and $\mathcal{N}$ are scalars, $\mathcal{J}^\mu$ and $\mathcal{Q^\mu}$ are vectors orthogonal to $u^\mu$ ($\mathcal{J}^\mu u_\mu=\mathcal{Q}^\mu u_\mu=0$), and $\mathcal{T}^{\mu\nu}$ is a symmetric, traceless and transverse tensor.
These quantities are defined as
\begin{eqnarray}
	\mathcal{E}&=&u_\mu u_\nu T^{\mu\nu}\,,		\qquad	\mathcal{P}=\frac{1}{d}\Delta_{\mu\nu}T^{\mu\nu}\,,	\qquad	\mathcal{N}=-u_\mu J^\mu\,, \qquad \mathcal{Q}_\mu=-\Delta_{\mu\lambda}u_\sigma T^{\lambda\sigma}\,,\nonumber\\
    \mathcal{J}_\mu&=&\Delta_{\mu\nu}J^\nu\,, \qquad
	\mathcal{T}_{\mu\nu}=\frac{1}{2}\left(\Delta_{\mu\lambda}\Delta_{\nu\sigma}+\Delta_{\nu\lambda}\Delta_{\mu\sigma}-\frac{2}{d}\Delta_{\mu\nu}\Delta_{\lambda\sigma}\right)T^{\lambda\sigma}\,.
 \label{eqn:coefficients_decomposition}
\end{eqnarray}
Note that, this decomposition is an identity that holds for all symmetric tensors and vectors. The hydrodynamic assumption comes into play when we express $\mathcal{E}$, $\mathcal{P}$, $\mathcal{N}$, $\mathcal{Q}^\mu$, $\mathcal{J}^\mu$, and $\mathcal{T}^{\mu\nu}$
of \eqref{eqn:tensor_current_standard_decomposition} as functions of hydrodynamic variables $\mu$, $T$, and $u^\mu$. We adopt the approach of a derivative expansion, typical in effective field theory, by assuming that the hydrodynamic fields deviate from equilibrium only on scales significantly larger than the microscopic mean free path, which serves as a UV cutoff. This means that first-order corrections are more significant than second-order ones, second-order terms outweigh third-order ones, and so forth. Formally, one writes
\begin{subequations}\label{eqn:gradient_expansion}
\begin{align}
	T^{\mu\nu}&=T^{\mu\nu}_{(0)}+T^{\mu\nu}_{(1)}+T^{\mu\nu}_{(2)}+\dots+T^{\mu\nu}_{(n)}+\mathcal{O}(\partial^{(n+1)})\,,\\
	J^\mu&=J^\mu_{(0)}+J^\mu_{(1)}+J^\mu_{(2)}+\dots+J^\mu_{(n)}+\mathcal{O}(\partial^{(n+1)})\,.
\end{align}
\end{subequations}
The perfect fluid contains zero derivatives, while Navier-Stokes equations has first order derivative corrections~\footnote{We would like to remark that gradient expansions are not exclusive to fluids; they
also apply to systems like solids or quasi-electrons in a lattice, so the presence of a gradient expansion doesn't imply the system is a fluid.}.
\subsection{Perfect fluid}\label{sec:ideal_fluid}
Since all transverse vectors and transverse traceless tensors we can construct with $\mu$, $T$, and $u^\mu$ and their derivatives are at least first-order in gradients, it follows that the ideal fluid will have $\mathcal{Q}^\mu = \mathcal{J}^\mu = \mathcal{T}^{\mu\nu} = 0$. Thus, we need to write $\mathcal{E}$, $\mathcal{P}$, and $\mathcal{N}$ in terms of the zeroth-order scalars $\mu$ and $T$. 

To advance, we note that in global thermodynamic equilibrium (in the fluid rest frame), the stress-energy tensor and the current assume a universal form, specifically $T^{\mu\nu}_\text{RF}=\text{diag}(\epsilon, P, \dots, P)$ and $J^\mu_\text{RF}=(n, \vect{0})$. These expressions define the equilibrium energy density $\epsilon$, pressure $P$, and charge density $n$. 
Moreover, performing a Lorentz boost will lead to the covariant expressions for a generic fluid moving with velocity $u^\mu$, hence, the final expressions are 
\begin{subequations}\label{eqn:ideal_fluid_constitutive_relations}
\begin{align}
	T^{\mu\nu}&=\epsilon u^\mu u^\nu + P\Delta^{\mu\nu}\,,\\
	J^\mu&=n u^\mu\,.
\end{align}
\end{subequations}
By design, these equations are valid in global thermodynamic equilibrium. However, perfect hydrodynamics assumes that the same form applies when the thermodynamic variables are treated as slowly varying fields in local thermal equilibrium. Thus, for an ideal fluid, we have $\mathcal{E}(x) = \epsilon(x)$, $\mathcal{P}(x) = P(x)$, and $\mathcal{N}(x) = n(x)$.

It is evident from this discussion that an equilibrium equation of state $P(T, \mu)$ is also necessary to determine the explicit constitutive relations in terms of the fundamental hydrodynamic fields. Therefore, we assume the existence of a suitable equation of state and, using the thermodynamic relations, we have~\cite{Landau:StatisticalPhysicsVolume}
\begin{subequations}\label{eqn:thermodynamic_identities}
\begin{align}
	\dif P&=\frac{\partial P}{\partial T}\biggr\rvert_\mu\dif T+\frac{\partial P}{\partial\mu}\biggr\rvert_T\dif\mu=s\dif T+n\dif\mu\,,\\
	\epsilon&=-P+Ts+\mu n \,,
\end{align}
\end{subequations}
where $s$ is the entropy density.

If we combine the longitudinal component of \eqref{eqn:stress_energy_conservation}, namely $u_\nu \partial_\mu T^{\mu\nu}=0$, with the equation~\eqref{eqn:current_conservation} and keeping only the zeroth-order in derivatives,
we obtain the conservation equation for entropy density
\begin{equation}
	\partial_\mu \left(s u^\mu\right)=0\,,
\end{equation}
which is interpreted as the perfect fluid entropy current and is conserved locally. Hence, perfect fluid hydrodynamics is non-dissipative.
\subsection{Frame choice}\label{sec:frame_choice}
Before discussing the first-order corrections to the perfect fluid, which lead to the relativistic Navier-Stokes equations, we must address a common issue in effective field theory constructions. This issue involves the redefinition of fundamental fields needed to perform the gradient expansion.

At the first order in derivatives, we can redefine the hydrodynamic variables in terms of first-order quantities that disappear in equilibrium. This ensures that various definitions coincide in global thermodynamic equilibrium when gradients are absent. A particular definition of the fields $T$, $\mu$, and $u^\mu$ is often referred to as a \emph{frame choice}. This is due to the lack of a microscopic definition of the hydrodynamic fields in out-of-equilibrium scenarios. For example, there is no temperature operator whose expectation value would give $T(x)$ for states that are not in equilibrium. The appropriate way to conceptualize the constitutive relations is to regard $T(x)$, $\mu(x)$, and $u^\mu(x)$ as auxiliary parameters used to parametrize $T^{\mu\nu}$ and $J^\mu$, which do have a microscopic out-of-equilibrium definition. A frame change may modify the hydrodynamic fields but will not alter the expectation value of the stress-energy tensor and the current, only their appearance in terms of the local variables.

From an effective field theory viewpoint, it seems natural to have the flexibility to redefine fields that differ by derivative quantities. This freedom is inherent when discussing the EFT approach for dissipative systems \cite{Glorioso:LecturesNonequilibriumEffective,Endlich:DissipationEffectiveField}. Since frame choices are essentially redundant within the theory, they can be viewed as gauge-like transformations \cite{Dore:FluctuatingRelativisticDissipative}. Furthermore, much like gauge theories, they can be formulated in a gauge-invariant manner, meaning that it is feasible to construct a theory of hydrodynamics without imposing a specific frame, relying solely on frame-invariant scalars and vectors \cite{Bhattacharya:TheoryFirstOrder}.

This depicts that the coefficients $\mathcal{E}$, $\mathcal{P}$, and $\mathcal{N}$ take the general form as
\begin{subequations}
\begin{align}
	\mathcal{E}&=\epsilon + \delta \mathcal{E}=\epsilon(T,\mu)+f_\mathcal{E}(\partial T, \partial\mu, \partial u)\,,\\
	\mathcal{P}&=P + \delta \mathcal{P}=P(T,\mu)+f_\mathcal{P}(\partial T, \partial\mu, \partial u)\,,\\
	\mathcal{N}&=n + \delta \mathcal{N}=n(T,\mu)+f_\mathcal{N}(\partial T, \partial\mu, \partial u)\,,
\end{align}
\end{subequations}
where $\delta \mathcal{E}$, $\delta \mathcal{P}$, and $\delta \mathcal{N}$ are higher-order corrections involving derivatives of the hydrodynamic fields $T$, $\mu$, and $u^\mu$
and are determined by our selection of the out-of-equilibrium definitions of local temperature, chemical potential, and fluid velocity.

Let us now consider a frame transformation with $\delta$ being the first-order in gradient correction
\begin{subequations}\label{eqn:frame_transformation}
\begin{align}
	T(x)	&\longrightarrow\quad	 T'(x)=T(x)+\delta T(x)\,,\\
	\mu(x)	&\longrightarrow\quad 	\mu'(x)=\mu(x)+\delta\mu(x)\,,\\
	u^\mu(x)&\longrightarrow\quad 	u'^\mu(x)=u^\mu(x)+\delta u^\mu(x)\,.
\end{align}
\end{subequations}
Note that $\delta u^\mu$ is a transverse vector, satisfying $u \cdot \delta u = 0$. This condition is necessary to preserve the normalization of the $u^\mu$ at the given order in derivatives, specifically $u\cdot u = -1 + \mathcal{O}(\partial^2)$.

Using the redefined $u^\mu$, we can decompose the stress-energy tensor and the current as in \eqref{eqn:coefficients_decomposition}. It is important to remember that $T^{\mu\nu}$ and $J^\mu$ are invariant under frame choice, and $\mathcal{Q}^\mu$ and $\mathcal{J}^\mu$ are themselves first-order in derivatives. This allows us to determine how the coefficients of the decomposition change under a frame transformation. This yields
\begin{subequations}\label{eqn:frame_transformation_coefficients}
\begin{align}
\delta\mathcal{E}&=0\,,\qquad
\delta\mathcal{P}=0\,,\qquad
\delta\mathcal{N}=0\,,\\
	\delta\mathcal{Q}^\mu&=-(\mathcal{E}+\mathcal{P})\delta u^\mu\,,
 \qquad\delta\mathcal{J}^\mu=-\mathcal{N}\delta u^\mu\,,\\
	\delta\mathcal{T}^{\mu\nu}&=0\,.
\end{align}
\end{subequations}
From above relations, it is easy to notice that one can always choose $\delta u^\mu$ such that either $\mathcal{J}^\mu = 0$ or $\mathcal{Q}^\mu = 0$. The former case is often referred to as the Eckart frame, while the latter is known as the Landau frame. Consequently, in the Landau frame it is possible to have zero charge density, while still having a non-zero charge current. Furthermore, note that the Eckart frame is mostly used when the U(1) conserved charge represents the baryon number.

Although frames are not physical in the hydrodynamic regime—they are simply redefinitions of the fields used to characterize $T^{\mu\nu}$ and $J^\mu$—they still have physical meaning. Specifically, the Eckart frame is defined such that there is no charge flow in the (local) fluid rest frame, whereas we define the Landau frame assuming no energy current in the fluid rest frame. In principle, there are infinite number of frames; however, in practice, only a few are commonly used. 
Notably useful frames include the thermodynamic frame \cite{Jensen:HydrodynamicsEntropyCurrent} (to be discussed later), BDNK frame \cite{Kovtun:FirstorderRelativisticHydrodynamics,Bemfica:NonlinearCausalityGeneral} and the recently introduced density frame \cite{Armas:2020mpr,Basar:2024qxd}.

The first equation in \eqref{eqn:frame_transformation_coefficients} indicates that $$\epsilon(T, \mu) + f_\mathcal{E}(\partial T, \partial \mu, \partial u) = \epsilon(T', \mu') + f'_\mathcal{E}(\partial T', \partial \mu', \partial u').$$ In analogy, for $\mathcal{N}$ and $\mathcal{P}$, we have
\begin{subequations}
\begin{align}
	f'_\mathcal{E}&=f_\mathcal{E}-\left(\frac{\partial\epsilon}{\partial T}\right)_\mu\delta T-\left(\frac{\partial\epsilon}{\partial\mu}\right)_T\delta\mu \,,\\
	f'_\mathcal{P}&=f_\mathcal{P}-\left(\frac{\partial p}{\partial T}\right)_\mu\delta T-\left(\frac{\partial p}{\partial\mu}\right)_T\delta\mu \,,\\
	f'_\mathcal{N}&=f_\mathcal{N}-\left(\frac{\partial n}{\partial T}\right)_\mu\delta T-\left(\frac{\partial n}{\partial\mu}\right)_T\delta\mu \,.
\end{align}
\end{subequations}
We can always choose the out-of-equilibrium definition of $\mu$ and $T$ which make two of the three functions $f$ vanish. Commonly, $f'_\mathcal{E}$ and $f'_\mathcal{N}$ are set to zero, ensuring that $\mathcal{E} = \epsilon$ and $\mathcal{N} = n$ at all orders in derivatives.
\subsection{Navier-Stokes}\label{sec:order_one}
\begin{figure}[t]
	\centering
	\includegraphics[width=0.4\textwidth]{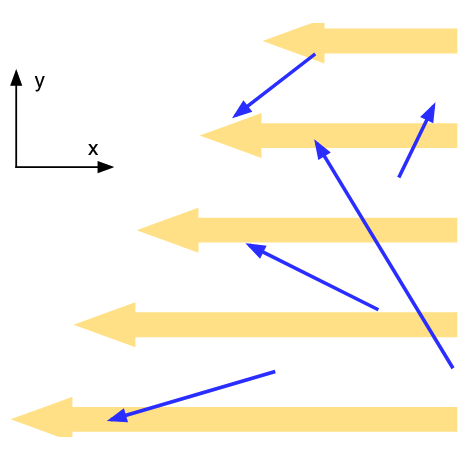}
	\caption{A stationary flow of a perfect fluid with a velocity gradient is depicted. Thin arrows represent particles transferring $x$-momentum in the $y$-direction, eventually equilibrating the inhomogeneous velocity profile~\cite{Kovtun:LecturesHydrodynamicFluctuations}.}
	\label{fig::impossible_flow}
\end{figure}

The model of ideal fluid dynamics  predicts unphysical solutions meaning that considering the constitutive relations truncated at zero order in derivatives, the resulting solutions to the hydrodynamic conservation equations describe unphysical flows. 
For instance, a solution to the ideal hydrodynamic equations such as
an equilibrated fluid moving in the $x$ direction with its gradient along $y$ is unphysical.
In reality, fluid particles transfer momentum between layers, making the flow homogeneous, as illustrated in Figure~\ref{fig::impossible_flow}. Therefore, it is essential to study the first-order dissipative corrections, where such flows are not equilibrium solutions.
Assuming Landau frame, we will specify the out-of-equilibrium definition of $u^\mu$ that follows $\mathcal{Q}^\mu = 0$. 
Moreover, we  fix $\mu$ and $T$ such that $\mathcal{E} = \epsilon$ and $\mathcal{N} = n$ with the Landau frame matching conditions
\begin{equation}\label{eqn:landau_matching_conditions}
	u_\mu T^{\mu\nu}=-\epsilon u^\nu \,,\qquad u_\mu J^\mu=-n \,,
\end{equation}
such that the fluid velocity four-vector becomes an eigenvector of the stress-energy tensor with energy as its eigenvalue.

We now express the free coefficients $\mathcal{P}$, $\mathcal{J}^\mu$, and $\mathcal{T}^{\mu\nu}$ (those not fixed by the frame choice) in terms of the gradients of the fundamental fields. At this order, there are only three one-derivative scalars: the material derivatives of temperature and chemical potential, $u^\alpha \partial_\alpha T$, $u^\alpha \partial_\alpha \mu$, and the expansion scalar $\theta = \partial_\alpha u^\alpha$.

Additionally, there are three transverse vectors: $\Delta^{\mu\nu} \partial_\nu T$, $\Delta^{\mu\nu} \partial_\nu \mu$, and $\Delta^{\mu\nu} a_\nu$, where $a_\mu = u^\alpha \partial_\alpha u_\mu$ represents the acceleration. Lastly, there is only one traceless symmetric first-order tensor we can construct, the shear tensor:
\begin{equation}\label{eqn:shear_tensor}
\sigma^{\mu\nu}=\Delta^{\mu\alpha}\Delta^{\nu\beta}\left(\partial_\alpha u_\beta+\partial_\beta u_\alpha-\frac{2}{d}\eta_{\alpha\beta}\,\theta \right) \,.
\end{equation}
Focusing on the scalar sector, we have already specified $\mathcal{E}$ and $\mathcal{N}$ using the Landau matching conditions, so we are left with $\mathcal{P}$. This scalar $\mathcal{P}$ should be expressed as the most general combination of the zero-order and first-order scalars mentioned above,
\begin{equation}
\mathcal{P}=P+c_1u^\alpha\partial_\alpha T+c_2u^\alpha\partial_\alpha\mu+c_3\theta+\mathcal{O}(\partial^2) \,.
\end{equation}
Since hydrodynamics is a formal series expansion, therefore, by utilizing the ideal fluid constitutive relations \eqref{eqn:ideal_fluid_constitutive_relations} in  $\partial_\mu J^\mu = 0$ and $u_\mu \partial_\nu T^{\mu\nu} = 0$, we obtain two equations that connect the three first-order scalars. This implies that, for instance, $u^\alpha \partial_\alpha T$ is proportional to $u^\alpha \partial_\alpha \mu$, up to higher order corrections in derivatives $\mathcal{O}(\partial^2)$. The same relationship holds for the expansion. Thus, at first order in derivatives, there is only one independent scalar and one generally chooses $\partial_\mu u^\mu$. 

Consequently, we can finally express $\mathcal{P}$ as the sum of the zeroth-order term and the first-order correction:
\begin{equation}
	\mathcal{P}=P-\zeta\partial_\mu u^\mu+\mathcal{O}(\partial^2) \,,
\end{equation}
where $\zeta=\zeta(T,\mu)$ is the bulk viscosity which is a transport coefficient.

We now focus on the remaining unspecified transverse vector, $\mathcal{J}^\mu$. As with previous cases, at first order in derivatives, there are three transverse vectors. In addition, we have one transverse lowest-order equation of motion, $\Delta_{\mu\nu} \partial_\alpha T^{\alpha\nu} = 0$. This connects the three first-order vectors up to higher-order corrections, implying that there are only two independent terms. Thus, we can express $\mathcal{J}^\mu$ as
\begin{equation}\label{eqn:dissipative_current}
\mathcal{J}^\mu=-\sigma T\Delta^{\mu\nu}\partial_\nu\frac{\mu}{T}+\chi_T\Delta^{\mu\nu}\partial_\nu T+\mathcal{O}(\partial^2) \,,
\end{equation}
where we encounter two more transport coefficients, charge conductivity $\sigma$ and $\chi_T$.

Notice that the only transverse symmetric traceless tensor is the shear tensor \eqref{eqn:shear_tensor} and it is proportional to $\mathcal{T}^{\mu\nu}$
\begin{equation}
	\mathcal{T}^{\mu\nu}=-\eta \sigma^{\mu\nu}+\mathcal{O}(\partial^2) \,,
\end{equation}
where $\eta$ is the transport coefficient namely shear viscosity.

If we had worked with the Eckart frame, we would have obtained the same equations for $\mathcal{T}^{\mu\nu}$ and $\mathcal{P}$, while what we wrote for $\mathcal{J}^\mu$ would now apply to $\mathcal{Q}^\mu$. 
It's worth noting that hydrodynamics has two ambiguities: one concerns the frame choice (such as Eckart or Landau), and the other involves selecting the basis of independent coefficients used to formulate the hydrodynamics constitutive relations. For instance, in our case, we chose not to include the acceleration when writing the constitutive relation for $\mathcal{J}^\mu$, whereas in the Eckart frame, one, generally, includes the acceleration term instead of $\Delta^{\mu\nu}\partial_\nu(\mu/T)$. This choice might seem trivial, but it is crucial to ensure that computed quantities do not depend on these ambiguities by working at the correct derivative order \cite{Kovtun:LecturesHydrodynamicFluctuations}.

Regarding transport coefficients: these quantities cannot be determined only using hydrodynamics, as they also depend on the microscopic dynamics, whereas hydrodynamics only describes the macroscopic dynamics in the infrared (IR). While for certain transport coefficients related to quantum anomalies, hydrodynamics alone can provide information \cite{Son:HydrodynamicsTriangleAnomalies}, in general, transport coefficients are unknown functions of the scalar fields.

Kubo formulae provide a method to compute these coefficients from a specific microscopic model, at least in principle. However, from the perspective of hydrodynamics, they are simply unknown functions of the scalar fields.

The Kubo formulae do not depend on the frame choices and can be employed to match the microscopic theory to the hydrodynamics EFT. Consequently, the transport coefficients are frame-invariant: while they may appear in different positions in the constitutive relations depending on the chosen frame, their form in terms of the microscopic theory remains unique and unaffected by this choice.
\subsection{Entropy current}\label{sec:entropy_current}
The hydrodynamic constitutive relations derived above are based solely on Lorentz covariance and derivative expansion. However, hydrodynamics necessitates additional conditions. Particularly, when formulating hydrodynamics, one imposes a local form of the thermodynamic second law, requiring that the entropy production is locally positive. In the following, we will shed light on another type of constraint stemming from the discrete symmetries of the microscopic theory.

In equilibrium, we can express the entropy current $S^\mu = su^\mu$ for constant velocity, and we have demonstrated that entropy is not produced for the case of the ideal fluid, i.e., $\partial_\mu (s u^\mu) = 0$. 
In first-order hydrodynamics, we posit that $S^\mu$ receives derivative corrections, aligning with the covariantization of thermodynamics. Specifically, we start with the Euler relation $Ts = P + \epsilon - \mu n$ and reformulate it in a covariant manner~\cite{Eckart:ThermodynamicsIrreversibleProcesses,Israel:ThermodynamicsRelativisticSystems}
\begin{equation}
	TS^\mu=Pu^\mu-T^{\mu\nu}u_\nu-\mu J^\mu  \,,
\end{equation}
where using Eq.~\eqref{eqn:tensor_current_standard_decomposition}, we obtain the expression for the entropy current
\begin{eqnarray}\label{eqn:canonical_entropy_current}
	S^\mu &=& S^\mu_{(0)} u^\mu+\frac{1}{T}\mathcal{Q}^\mu-\frac{\mu}{T}\mathcal{J}^\mu \,,\nonumber\\
 &=&\left[s+\frac{1}{T}\left(\mathcal{E}-\epsilon\right)-\frac{\mu}{T}\left(\mathcal{N}-n\right)\right]u^\mu+\frac{1}{T}\mathcal{Q}^\mu-\frac{\mu}{T}\mathcal{J}^\mu \,,
\end{eqnarray}
where $S^\mu_{(0)}$ is the entropy density in the rest frame whereas the remaining terms are the entropy flow in the rest frame.

This is known as the canonical entropy current, and it is frame invariant. While the form of the current in terms of the gradients may depend on the frame definition, the value of the current remains constant. There are other contributions to the entropy current, referred to as non-canonical, which play significant roles when discussing hydrostatic or non-dissipative fluids.

In the Landau frame, the expression of entropy current is~\cite{Hartnoll:TheoryNernstEffect,Kovtun:LecturesHydrodynamicFluctuations,Landau:FluidMechanicsVolume}
\begin{equation}
	S^\mu=su^\mu-\frac{\mu}{T}\mathcal{J}^\nu \,.
\end{equation}
We use the above expression with the Landau-frame constitutive relations to enforce local second law of thermodynamics
\begin{equation}
	\partial_\mu S^\mu\geq0  \,,
\end{equation}
on the flows that might be the solutions to~\eqref{eqn:stress_energy_conservation}--\eqref{eqn:current_conservation}.
Using the Gibbs–Duhem equation $$\partial P=s\partial T+n\partial\mu\,,$$ one gets
\begin{align}
	\partial_\mu S^\mu&=\frac{\zeta}{T}\theta^2+\frac{\eta}{T}\sigma_{\mu\nu}\sigma^{\mu\nu}+\sigma\left(T\Delta^{\mu\nu}\partial_\nu\frac{\mu}{T}\right)^2-T\chi_T\Delta^{\mu\nu}\partial_\nu\frac{\mu}{T}\partial_\mu T\geq0  \,,
\end{align}
where $\theta$ is the expansion scalar and $\sigma_{\mu\nu}$ is the shear tensor.
Since the right hand side of the equation must be positive definite we find the following constraints on the transport coefficients
\begin{equation}\label{eqn:entropy_constraints}
	\eta\geq0 \,,\qquad \zeta\geq0 \,,\qquad \sigma\geq0 \,,\qquad \chi_T=0 \,.
\end{equation}
Even though Lorentz covariance requires us to have four transport coefficients for first order relativistic hydrodynamics, the additional constraint on the positivity of entropy production reduces this to three.

Let us present here the complete first-order Landau frame constitutive relations for a normal fluid, derived from the requirements of Lorentz covariance, gradient expansion, and entropy production positivity
\begin{subequations}\label{eqn:landau_constitutive_relations}
\begin{align}
	T^{\mu\nu}&=\epsilon u^\mu u^\nu+P\Delta^{\mu\nu}-\eta\sigma^{\mu\nu}-\zeta \theta \Delta^{\mu\nu}+\mathcal{O}(\partial^2) \,,\\
	J^\mu&=n u^\mu-\sigma T\Delta^{\mu\nu}\partial_\nu\frac{\mu}{T}+\mathcal{O}(\partial^2) \,.
\end{align}
\end{subequations}
Certainly, an alternative method to derive the same constitutive relations is to select a fixed frame and directly calculate the divergence of the canonical entropy current. Analyzing the resulting expression allows for the derivation of the constitutive relations.

\subsection{Hydrostatic generating functional}\label{sec:generating_functional}
The hydrostatic generating functional, as described in \cite{Jensen:HydrodynamicsEntropyCurrent,Banerjee:ConstraintsFluidDynamics}, is a potent technique for computing $n$-point functions, and consequently, the constitutive relations, from a path-integral approach in the hydrostatic regime. This method is applicable when variables are time-independent and, in the absence of external sources, constant.

Hydrostatic flow, along with the dissipative sector described above, represents two of the eight possible classes of fluid transport categorized in \cite{Haehl:AdiabaticHydrodynamicsEightfold,Haehl:EightfoldWayDissipation} using the concept of adiabatic flows.

Let us consider a charged fluid in the presence of spacetime and gauge sources and assume it is in time-independent hydrostatic equilibrium. This is possible when a time-like Killing field $V^\mu$ exists, parametrizing time-translations that allow all thermodynamic quantities and given external sources to have vanishing Lie derivative, $\mathcal{L}_V$, with respect to $V^\mu$. 
Physically, this implies that the sources are slowly varying functions, and the time derivative in the frame comoving with $V^\mu$, when $V^\mu = (1, 0,0,0)$, is vanishing~\footnote{We remark that the Lie derivative's ability to describe the invariance of tensor fields under the flow of symmetries, its coordinate independence, and its compatibility with the geometry of spacetime make it the ideal tool for generalizing mechanical equilibrium in general relativity. This allows equilibrium concepts to be seamlessly extended from flat spacetime to the curved, dynamic spacetimes of general relativity.}

Since we are handling equilibrium configurations in the presence of adiabatically turned-on sources, we can compute the Green functions utilizing the generating functional by differentiating with respect to the metric and gauge sources: $g_{\mu\nu}$ and $A_\mu$.

In the frame where Killing field is constant consider the collection of zero-frequency $n$-point functions truncated up to $m$-th order in powers of $\mathbf{k}$. These can be inverse-Fourier transformed to yield a series of correlators in position space, which hold approximately on scales significantly larger than the microscopic system's correlation. After integrating this series of approximate Green functions, we obtain the equilibrium generating functional for the truncated correlators that depend on the sources
\begin{equation}
	W_m[g,A]=\int\dif^dx\ \mathcal{W}[\text{sources}(x)] \,,
\end{equation}
where $\mathcal{W}$ consists terms up to $m$-order in derivatives. Since this expression is covariant, it is valid also for non-comoving frames with Killing vector.

For $W_m$ to reproduce Eq.~\eqref{eqn:conservation_equation_sources}, it should be diffeomorphism and gauge invariant. Therefore, $\mathcal{W}$ needs to depend on diffeomorphism and gauge invariant scalars. These scalar observables must be spacetime dependent but can be non-local in Euclidean time to describe systems at finite temperature.

Consider two quantities, one, an invariant length $L$ of the time circle in the Euclidean theory and, two, the Polyakov loop $P_A$ of $\mathrm{U(1)}$ gauge field in the same direction. $L$ can be evaluated as
\begin{equation}
L=\int_0^\beta\dif\tau\sqrt{g_{\tau\tau}} \,,
\end{equation}
in comoving frame with $V^\mu$, with $\beta = 1/T_0$ being the coordinate period. After making the expression covariant and rotating back to non-imaginary time, we obtain
\begin{equation}
	L=\beta\sqrt{-V^2} \,,\qquad \ln P_A=\beta V^\mu A_\mu \,,
\end{equation}
We are in a position to obtain temperature, chemical potential and fluid velocity from these quantities
\begin{subequations}\label{eqn:thermodynamic_frame_definitions}
	\begin{align}
		T&=\frac{1}{L}=\frac{T_0}{\sqrt{-V^2}} \,,\\
		\mu&=\frac{\ln P_A}{L}=\frac{V^\mu A_\mu-\Lambda_\beta}{\sqrt{-V^2}} \,,\\
		u^\mu&=\frac{V^\mu}{\sqrt{-V^2}} \,.
	\end{align}
\end{subequations}
To preserve gauge invariance a gauge parameter, $\Lambda_\beta$, is included. 
For a given $m$-order derivative, the theory contains $N_m$ scalar quantities $s_{m,1}, s_{m,2}, \dots, s_{m,N_m}$, which are computed from the invariants \eqref{eqn:thermodynamic_frame_definitions} and their derivatives. Consequently, the most general hydrostatic generating functional at $m$-order is expressed as
\begin{equation}
\label{eqn:generating_functional}
	W_m[g,A]=\int\dif^dx\sqrt{-g}\left[P(s_0)+\sum_{n=1}^m\sum_{i=1}^{N_n} F_{n,i}(s_0)s_{n,i}\right] \,,
\end{equation}
where $s_0$ represents the zeroth-order scalars, which for a simple fluid are $s_0 = \{T, \mu\}$. The functions $F_{n,i}$ are unknown functions of these zeroth-order scalars, and $P$ can be interpreted as the pressure. In the absence of sources, when all derivative corrections to $W_m$ are zero, $W$ matches with the free energy.

With this expression, we can now determine the hydrostatic constitutive relations for $T^{\mu\nu}$ and $J^\mu$, which are the one-point functions, by differentiating $W_m$ with respect to $g_{\mu\nu}$ and $A_\mu$ respectively
\begin{equation}
	\langle T^{\mu\nu}\rangle=\frac{2}{\sqrt{-g}}\frac{\delta W_m}{\delta g_{\mu\nu}} \,,\qquad \langle J^\mu\rangle=\frac{1}{\sqrt{-g}}\frac{\delta W_m}{\delta A_\mu} \,.
\end{equation}
The constitutive relations derived from the hydrostatic generating functional manifest in a specific hydrodynamic frame known as the thermodynamic frame. Unlike the Landau frame, \eqref{eqn:landau_matching_conditions}, the thermodynamic frame implies that the definitions of temperature, chemical potential, and fluid velocity are maintained without modification on shell, relative to their equilibrium values \eqref{eqn:thermodynamic_frame_definitions}. It's important to clarify that while the term `thermodynamic frame' is commonly used, it does not denote a true frame but rather a category of frames. This is because the thermodynamic frame prescribes how to handle the hydrostatic sector of the constitutive relations, but it does not address the dissipative components, which require additional frame-fixing conditions.

One might also inquire about the fate of the entropy current within the hydrostatic sector. The hydrostatic solutions are inherently non-dissipative, thus ensuring $\partial_\mu S^\mu = 0$; however, achieving this outcome requires a modification of the entropy current itself compared to its canonical form. The total entropy current is
\begin{equation}
	S^\mu = S^\mu_\text{can}+S^\mu_\text{non-can} \,,
\end{equation}
where the first and second terms are the canonical and non-canonical entropy currents, respectively.
The latter is to ensure that the entropy is not produced due to hydrostatic derivative corrections~\cite{Bhattacharyya:EntropyCurrentPartition,Bhattacharyya:EntropyCurrentEquilibrium}.
\subsubsection{Hydrostatic conditions}\label{sec:hydrostatic_constraints}
For the fluid to be in hydrostatic flows, its hydrodynamic variables must satisfy specific conditions that essentially define equilibrium. These constraints are derived by demanding that the Lie derivatives of the sources and the thermodynamic quantities \eqref{eqn:thermodynamic_frame_definitions} with respect to $V^\mu$ are zero. To move forward, we first decompose the $\mathrm{U(1)}$ electromagnetic field strength tensor by projecting parallel to the fluid flow velocity
\begin{equation}
E^\mu = F^{\mu\nu} u_\nu, \quad B^\mu = \frac{1}{2} \epsilon^{\mu\nu\rho\sigma} u_\nu F_{\rho\sigma}\,,
\end{equation}
where $E^\mu$ is the electric field and $B^\mu$ is the magnetic field as seen by the fluid. These projections allow us to express the field strength tensor $F_{\mu\nu}$ in terms of $E^\mu$ and $B^\mu$ as follows:
\begin{equation}\label{eqn:electric_magnetic_decomposition}
F^{\mu\nu} = u^\mu E^\nu - u^\nu E^\mu + \epsilon^{\mu\nu\rho\sigma} u_\rho B_\sigma\,.
\end{equation}
This decomposition is useful for analyzing the equilibrium conditions in the presence of an electromagnetic field, as it separates the contributions parallel and perpendicular to the fluid flow.

Demanding $\mathcal{L}_V T = \mathcal{L}_V \mu = 0$ leads to the first two hydrostatic constraints \cite{Jensen:HydrodynamicsEntropyCurrent}. These constraints ensure that the temperature $T$ and chemical potential $\mu$ are constant along the flow defined by the Killing vector $V^\mu$. Formally, this can be expressed as:
\begin{equation}
V^\mu \partial_\mu T = 0, \quad V^\mu \partial_\mu \mu = 0\,.
\end{equation}
These conditions imply that both the temperature and the chemical potential remain unchanged in the direction of $V^\mu$, which characterizes the equilibrium flow of the fluid
\begin{eqnarray}\label{eqn:hydrostatic_constraints_relativistic}
		\partial_\mu T&=&-Ta_\mu \,,\\
		\partial_\mu\mu&=&E_\mu-\mu a_\mu \,,
	\end{eqnarray}
where $a^\mu=u \cdot \nabla u^\mu$ is the acceleration. Note that, the above equations
can be directly derived from the definitions in \eqref{eqn:thermodynamic_frame_definitions} and the properties of Lie derivatives.

The other constraints can be obtained from the Lie derivative of the fluid flow velocity, for which we notice that the fluid which is in equilibrium should have zero expansion and shear tensor. Specifically, requiring that $\mathcal{L}_V u^\mu = 0$ implies that the fluid velocity $u^\mu$ is aligned with the Killing vector $V^\mu$. 

In formal terms, this means that the expansion $\theta$ and the shear tensor $\sigma^{\mu\nu}$ must both vanish:
\begin{eqnarray}
\theta = \nabla_\mu u^\mu = 0\,,
\end{eqnarray}
\begin{eqnarray}
\sigma^{\mu\nu} = \Delta^{\mu\alpha} \Delta^{\nu\beta} \left( \nabla_{\alpha} u_{\beta} + \nabla_{\beta} u_{\alpha} - \frac{2}{d} \eta_{\alpha\beta} \nabla_\gamma u^\gamma \right) = 0\,.
\end{eqnarray}
These conditions ensure that the fluid flow is purely along the direction of $V^\mu$ without any expansion or distortion, which is characteristic of a fluid in hydrostatic equilibrium.
Finally, we have Bianchi identity of the gauge field, $\varepsilon^{\mu\nu\alpha\beta}\nabla_\nu F_{\alpha\beta}=0$.

These are equality-type constraints that the fluid variables must satisfy to maintain equilibrium when external sources are present. Any dissipative correction to the constitutive relations must be viewed as deviations from these constraints. For instance, in equilibrium, the fluid must have vanishing expansion $\theta = 0$, consistent with the fact that bulk viscosity is a dissipative transport coefficient, and the fluid generates entropy when expansion is non-zero.

For the vector sector, the second law of thermodynamics indicates that the two constraints \eqref{eqn:hydrostatic_constraints_relativistic} lead to a single dissipative transverse vector
\begin{equation}\label{eqn:kovtun_condition}
	E^\mu - \Delta^{\mu\nu}\partial_\nu\frac{\mu}{T} = 0 \,,
\end{equation}
that arises in the constitutive relation of the current~\eqref{eqn:constitutive_relations_curved_space}.
Note that the reverse is not true: there are transport coefficients that are non-dissipative (they cancel out when calculating the positivity of entropy production), yet they disappear in hydrostatic flows. These are referred to as non-dissipative non-hydrostatic terms and can be derived from a generalization of the above generating functional~\cite{Haehl:AdiabaticHydrodynamicsEightfold}.
\section{Linear Response Theory}
\label{sec:linear_response_theory}
Hydrodynamics being a non-linear theory depicts that different initial conditions can give rise to different fluid dynamics evolution. This makes hydrodynamical non-linear equations notoriously difficult to solve.

The difficulty that arise in hydrodynamics are often related to extending the theory beyond its regime of validity. Being a low energy effective field theory, hydrodynamics is valid when $\omega$ and $\vect{k}$ are small. However, if one would like to use hydrodynamic equations as a system of non-linear partial differential equations with predictability, one would have to evaluate those equations at high values of $\vect{k}$, where hydrodynamics is expected to fail.

Focusing on linearized hydrodynamics provides a more controlled approach~\cite{Kurkela:2017xis,Ochsenfeld:2023wxz,Bajec:2024jez,Brants:2024wrx}. This entails trusting the theory within its valid domain: we examine a uniform global thermodynamic equilibrium state characterized by constant temperature $T = \text{const}$, chemical potential $\mu = \text{const}$, and in the rest frame of the fluid $u^\mu = (1,0,0,0)$. We study small fluctuations around this equilibrium, particularly focusing on the expansion involving small wave-vectors and frequencies, where hydrodynamics is expected to be applicable.
Linearized hydrodynamics remains informative in several respects:
\begin{itemize}
    \item It allows access to retarded correlators, which can be compared with two-point functions obtained from other methods (such as holography, kinetic theory, QFT, etc.), leading to Kubo formulae.
    \item These correlators provide physical insights into the transport properties of the system, such as thermoelectric conductivities.
    \item It enables the discovery of new constraints on the constitutive relations arising from microscopic discrete symmetries.
    \item It offers insights into the stability of the theory against small perturbations and fluctuations.
\end{itemize}
In the following, rather than using $T$, $\mu$, and $u^\mu$, we will work with their thermodynamic conjugate quantities as independent variables: specifically, energy, charge, and momentum density. The rationale behind this is that these quantities have a microscopic definition through the operators $\hat{T}^{0\mu}$ and $\hat{J}^0$, enabling us to examine their role in the Hamiltonian and derive the correlators linked to these charges.
\subsection{Martin-Kadanoff method}
\label{sec:martin_kadanoff}
Here, we will briefly recap the motivation behind linear response theory, which aims to study the response of a system in thermal equilibrium to small deviations~\cite{Kadanoff:HydrodynamicEquationsCorrelation}.
Let us consider a collection of hydrodynamic fields $\phi_a(t,\vect{x})$ which are microscopically well-defined. Then they are coupled to their weak, slowly-varying sources $\lambda_a(t,\vect{x})$. The sources are switched on at $t=-\infty$ and let increase adiabatically; at $t=0$, they are switched off, and the system is allowed to evolve freely. 
In linearized hydrodynamics, the classical fields $\phi_a$ are governed by a set of linear equations which, in momentum space, are expressed as:
\begin{equation}
\label{eqn:hydrodynamic_equations_schematic}	
\partial_t\phi_a(t,\vect{k})+M_{ab}(\vect{k})\phi_b(t,\vect{k})=0 \,,
\end{equation}
where $\phi_a$ are, strictly speaking, $\delta\phi_a$, which represents the deviation from equilibrium: $\delta\phi_a(t,\vect{x}) = \phi_a(t,\vect{x}) - \phi_a$.
Here, $M_{ab}$ is a matrix computed by the hydrodynamic conservation laws and the constitutive relations. This system of equations applies in the hydrodynamic regime, characterized by long wavelengths ($\vect{k} \rightarrow 0$) and small frequencies ($\omega \rightarrow 0$), governing the free evolution of the fields at $t > 0$. To analyze the solutions, we employ a Laplace transform in time
\begin{equation}
	\phi_a(z,\vect{k})=\int_0^{\infty}\dif t\ e^{izt}\phi_a(t,\vect{k}) \,,
\end{equation}
with $z$ lying on the upper half complex plane. This makes the integral converge and we find
\begin{equation}
	(-i z\delta_{ab}+M_{ab})\phi_b(z,\vect{k})=\phi_a^0(\vect{k}) \,,
\end{equation}
where $\phi^0_a(\vect{k})=\phi_a(t=0,\vect{k})$ is the initial value.
For small fluctuations one writes
\begin{equation}\label{eqn:static_susceptibility}
\phi_a^0(\vect{k}\rightarrow0)=\chi_{ab}\lambda_b^0(\vect{k}\rightarrow0)   \qquad\Longrightarrow\qquad \chi_{ab}=\left(\frac{\partial\phi_a}{\partial\lambda_b}\right) \,,
\end{equation}
with $\chi_{ab}$ being the static thermodynamic susceptibility. Using above definition we can solve the hydrodynamic equations in terms of the initial conditions
\begin{equation}\label{eqn:phi_from_linear_equation}
	\phi_a(z,\vect{k})=(K^{-1})_{ab}\chi_{bc}\lambda_c^0(\vect{k}) \,,
\end{equation}
where $K_{ab}=-iz\delta_{ab}+M_{ab}(\vect{k})$.

We now aim to connect the average values of the hydrodynamic fields to the system's correlation functions.
Suppose we perturb the fields using slowly-varying sources that are adiabatically switched on at $ t = -\infty $ and then turned off at $ t = 0 $. For instance, let's consider $ \lambda(t,\vect{x}) = e^{\varepsilon t}\lambda^0(\vect{x})\theta(-t) $, where $ \varepsilon > 0 $ is a small adiabatic parameter with $ \theta(t) $ being the step function. We then relate the system's evolution at $ t > 0 $ to eq.~\eqref{eqn:phi_from_linear_equation}. To achieve this, we introduce a perturbation to the Hamiltonian that couples the hydrodynamic fields with their respective sources
\begin{equation}
	\delta\hat H(t)=-\int \dif^d\vect{x}\ \lambda_a(t,\vect{x})\hat\phi_a(t,\vect{x}) \,.
\end{equation}
From time-dependent perturbation theory of first-order, we can determine the operator $\hat\phi_a(t,\vect{x})$ response in the Heisenberg picture. If the system possesses a time-independent Hamiltonian, and we introduce a small perturbation, the expected value of the observable is modified
\begin{equation}
\delta\langle\hat\phi_a(t,\vect{x})\rangle=-i\int_{-\infty}^t \dif t'\langle[\hat\phi_a(t,\vect{x}),\delta\hat H(t')]\rangle \,,
\end{equation}
where $\langle\dots\rangle$ represents the thermal average $\langle\hat O\rangle=\text{Tr}(\hat\rho\hat O)$, with $\hat\rho$ as the density matrix (in the grand canonical ensemble, the Heisenberg operators are defined as $\hat H'=\hat H-\mu \hat N$, where $\mu$ is the chemical potential serving as the Lagrange multiplier of the charge number operator $\hat N$). Combining these two equations yields
\begin{equation}\label{eqn:correlation_real_space}
	\delta\langle\hat\phi_a(t,\vect{x})\rangle=-\int_{-\infty}^{\infty}\dif t'\int\dif^d x'\ G^R_{ab}(t-t',\vect{x}-\vect{x}')\lambda_b(t,\vect{x}) \,,
\end{equation}
with $G^R_{ab}$, retarded response function, expressed as
\begin{equation}
	G^R_{ab}(t-t',\vect{x}-\vect{x}')=-i\theta(t-t')\langle[\hat\phi_a(t,\vect{x}),\hat\phi_b(t',\vect{x}')]\rangle \,.
\end{equation}
In Fourier space, the average value simplifies, and the convolution becomes
\begin{equation}
\delta\langle\hat\phi_a(\omega,\vect{k})\rangle=-G^R_{ab}(\omega,\vect{k})\lambda_b(\omega,\vect{k}) \,.
\end{equation}
We begin by evaluating the Fourier transform in space
\begin{equation}\label{eqn:phi_fourier}
	\langle\hat\phi_a(t,\vect{k})\rangle=-\int_{-\infty}^0\dif t' e^{\varepsilon t'}G^R_{ab}(t-t',\vect{k})\lambda_b^0(\vect{k}) \,.
\end{equation}
Having Fourier transformed the retarded function in time, we obtain
\begin{equation}
	G^R(t-t',\vect{k})=\int_{-\infty}^{\infty}\frac{\dif \omega}{2\pi}G^R(\omega,\vect{k})e^{-i\omega(t-t')} \,.
\end{equation}
Due of the step function in $G^R(t,\vect{k})$, this function is vanishing for $t < 0$. This depicts that $G^R(\omega,\vect{k})$ proved to be analytic in the upper half complex plane $\omega$ allowing us to analytically continue $G^R(\omega,\vect{k})$ to the whole complex plane. We then evaluate $t'$ integral leading to
\begin{equation}
	\langle\hat\phi_a(t,\vect{k})\rangle=-\lambda^0_b(\vect{k})\int\frac{\dif \omega}{2\pi}G^R_{ab}(\omega,\vect{k})\frac{e^{-i\omega t}}{i\omega+\varepsilon} \,,
\end{equation}
where $\varepsilon$ is required for the integral convergence. We now multiply both sides by $e^{izt}$ (ensuring $\text{Im}(z) > 0$ for convergence) and integrate over $t$ from $0$ to $\infty$ (effectively applying the Laplace transform as done previously to our linearized hydrodynamic equations)~\cite{Kovtun:LecturesHydrodynamicFluctuations}
\begin{equation}
	\langle\hat\phi_a(z,\vect{k})\rangle=-\lambda^0_b(\vect{k})\int\frac{\dif \omega}{2\pi}\frac{G^R_{ab}(\omega,\vect{k})}{(i\omega+\varepsilon)(i(\omega-z)+\varepsilon)} \,.
\end{equation}
As previously noted, $G^R(\omega)$ is analytic in the upper half-plane, enabling us to integrate by closing the contour with $\text{Im}(\omega) > 0$. Inside this contour, there are two poles: one at $\omega = i\varepsilon$ and another at $\omega = z + i\varepsilon$. According to the residue theorem, we obtain
\begin{equation}\label{eqn:martin_kadanoff_kubo_formulae}
	\langle\hat\phi_a(z,\vect{k})\rangle=-\lambda^0_b(\vect{k})\frac{G^R_{ab}(z,\vect{k})-G^R_{ab}(z=0,\vect{k})}{iz} \,,
\end{equation}
where the parameter $z = 0$ is understood to be slightly above the real axis. To determine $G^R(z=0,\vect{k})$, we examine equation \eqref{eqn:phi_fourier} evaluated at $t = 0$. The integral's argument represents the Laplace transform evaluated at $z = 0$ (slightly above the real axis, starting from $z = i\varepsilon$).
\begin{equation}
	\langle\hat\phi_a(t=0,\vect{k})\rangle=-\int_0^{\infty}\dif t'\ e^{-\varepsilon t'}G^R_{ab}(t',\vect{k})\lambda^0_b(\vect{k})=-G^R_{ab}(z=0,\vect{k})\lambda^0_b(\vect{k}) \,.
\end{equation}
From this equation, we deduce that in the long wavelength limit, $G^R(z=0,\vect{k})$ corresponds to the negative of the static susceptibility $\chi(\vect{k})$ defined earlier (see equation \eqref{eqn:static_susceptibility}), thus $G^R(z=0,\vect{k}) = -\chi(\vect{k})$.

We can now compare this result with our previous expression from linearized hydrodynamics, yielding
\begin{equation}
	-\frac{1}{iz}\bigl(G^R_{ab}(z,\vect{k})+\chi_{ab}(\vect{k})\bigr)\lambda^0_b(\vect{k})=(K^{-1})_{ac}\chi_{cb}\lambda^0_b(\vect{k}) \,,
\end{equation}
and we obtain for the retarded response function matrix as
\begin{equation}\label{eqn:retarded_green_functions}
G^R_{ab}(z,\vect{k})=-(\delta_{ac}+iz(K^{-1})_{ac})\chi_{cb} \,,
\end{equation}
which is analytic in the upper half-plane of the complex variable $z$, enabling us to extend $G^R(\omega,\vect{k})$ to the entire complex plane as the analytic continuation of $G^R(z,\vect{k})$ from the upper half-plane. Given the various analytic properties of $G^R$, it may also relate to symmetric and advanced Green functions~\cite{Kovtun:LecturesHydrodynamicFluctuations}. Therefore, it can be demonstrated that
\begin{equation}
	-\text{Im}\ G^R_{aa}(\omega,\vect{k})\geq 0    \qquad \text{for}\quad\omega\geq 0 \,,
\end{equation}
which, when applied to the hydrodynamic response functions, indicates that all the transport coefficients are positive-definite $\sigma\geq0$, $\zeta\geq0$, and $\eta\geq0$.
\subsection{Variational method}
\label{sec:variational_method}
In the previous section, we outlined the classical method of calculating the retarded Green functions from hydrodynamics using linear response theory. This approach introduces sources corresponding to the conserved charge densities derived from equilibrium thermodynamics, offering clarity from a physical standpoint. However, it has limitations, particularly in its inability to capture all correlators associated with the currents~\cite{Kovtun:LecturesHydrodynamicFluctuations,Romatschke:RelativisticFluidDynamics}.

Another approach to derive the same quantities from a field theory perspective involves introducing sources that directly couple to the covariant forms of $T^{\mu\nu}$ and $J^\mu$. In any generic theory, external sources such as a curved metric $g_{\mu\nu}$ and a gauge field $A_\mu$ can be added. Consequently, a generating functional $W[A,g]$ can be formally constructed, whose variations provide all the connected $n$-point functions of the theory.

By varying $W$ with respect to the $g^{\mu\nu}$ or $A_\mu$, one obtains the one-point functions corresponding to the current and stress-energy tensor
\begin{equation}
	\delta W[A,g]=\int\dif x\sqrt{-g}\left(T^{\mu\nu}\delta g_{\mu\nu}+J^\mu\delta A_\mu\right) \,.
\end{equation}
Moreover, when variations of the metric and gauge fields are linked to diffeomorphism or gauge symmetry, the invariance of $W[A,g]$ under these transformations yields the conservation equations of hydrodynamics (assuming no anomalies)
\begin{subequations}\label{eqn:conservation_equation_sources}
	\begin{align}
		\nabla_\mu T^{\mu\nu}&=F^{\nu\lambda}J_\lambda \,,\\
		\nabla_\mu J^\mu&=0 \,,
	\end{align}
\end{subequations}
which are the generalization of \eqref{eqn:stress_energy_conservation} and \eqref{eqn:current_conservation} when we take into account external curvature and electromagnetic fields.

Although we lack access to a microscopic generating functional for the complete dissipative theory in hydrodynamics, we can still formulate the generators of the retarded Green functions in the presence of background sources as
\begin{equation}\label{eqn:green_functions_generators}
	\mathcal{T}^{\mu\nu}[A,g]=\sqrt{-g}\langle\hat T^{\mu\nu}\rangle \,,\qquad \mathcal{J}^\mu[A,g]=\sqrt{-g}\langle\hat J^\mu\rangle \,,
\end{equation}
with $\langle\hat T^{\mu\nu}\rangle$ and $\langle\hat J^\mu\rangle$ being the solutions of the hydrodynamical equations of motion in the presence of $A_\mu$ and $g_{\mu\nu}$. The equilibrium correlators in the absence of sources can be calculated as
\begin{subequations}\label{eqn:green_functions_variational}
	\begin{align}
		G^R_{J^\mu J^\nu}(x)&=-\frac{\delta\mathcal{J}^\mu(x)}{\delta A_\nu(0)}\biggr\rvert_{A=h=0} \,,	&	G^R_{T^{\mu\nu} J^\sigma}(x)&=-\frac{\delta\mathcal{T}^{\mu\nu}(x)}{\delta A_\sigma(0)}\biggr\rvert_{A=h=0} \,,\\
		G^R_{J^\sigma T^{\mu\nu}}(x)&=-2\frac{\delta\mathcal{J}^\sigma(x)}{\delta h_{\mu\nu}(0)}\biggr\rvert_{A=h=0} \,,	&	G^R_{T^{\mu\nu} T^{\sigma\rho}}(x)&=-2\frac{\delta\mathcal{T}^{\mu\nu}(x)}{\delta h_{\rho\sigma}(0)}\biggr\rvert_{A=h=0} \,,
	\end{align}
\end{subequations}
where the metric fluctuation is $h_{\mu\nu}=g_{\mu\nu}-\eta_{\mu\nu}$.

The procedure to compute these quantities involves obtaining the constitutive relations in the Landau frame in the presence of background sources, at first order in derivatives, expressed as
\begin{align}\label{eqn:constitutive_relations_curved_space}
	T^{\mu\nu}&=\epsilon u^\mu u^\nu+P\Delta^{\mu\nu}\nonumber\\
	&\quad-2\eta\Delta^{\mu\alpha}\Delta^{\nu\beta}\left(\nabla_\alpha u_\beta+\nabla_\beta u_\alpha-\frac{2}{d}g_{\alpha\beta}\nabla_\lambda u^\lambda\right)-\zeta\Delta^{\mu\nu}\nabla_\lambda u^\lambda \,,\nonumber\\
	J^\mu&=n u^\mu+\sigma\left(E^\mu-T\Delta^{\mu\nu}\nabla_\nu\frac{\mu}{T}\right) \,,
\end{align}
where $E_\mu = F_{\mu\nu}u^\nu$ represents the covariant electric field, arising naturally from the entropy current in the presence of sources. Subsequently, we solve the hydrodynamic equations in Fourier space, linearizing around global thermodynamic equilibrium with constant temperature, chemical potential and fluid velocity in the presence of small fluctuating sources $\delta A_\mu$ and $\delta h_{\mu\nu}$. Inserting the resulting solution into the generators \eqref{eqn:green_functions_generators} and computing the functional derivatives \eqref{eqn:green_functions_variational} grants access to all the Green functions.
The Green functions acquired through this method might exhibit discrepancies compared to those obtained via the Martin-Kadanoff approach due to contact terms arising from the $\sqrt{-g}$ term in the generators \eqref{eqn:green_functions_generators}, as discussed in \cite{Herzog:LecturesHolographicSuperfluidity}.
\subsection{Discrete symmetries: Onsager relations}
\label{sec:onsager_relations}
The retarded functions must adhere to the symmetries, with time-reversal symmetry imposing stringent constraints on the transport coefficients. Consider a Hermitian field $\hat\phi_a(t,\vect{x})$ transforming under time reversal as $\hat\Theta\hat\phi_a(t,\vect{x})\hat\Theta^{-1}=\eta_a\hat\phi_a(-t,\vect{x})$, where $\hat\Theta$ represents the anti-unitary time-reversal operator and $\eta_a=\pm 1$ denotes the eigenvalue of $\hat\phi_a$. In a scenario where the microscopic system preserves time-reversal symmetry, i.e., $[H,\hat\Theta]=0$, the retarded response functions must satisfy
\begin{equation}
	G^R_{ab}(t,\vect{x})=G^R_{ba}(t,-\vect{x})\eta_a\eta_b \,.
\end{equation}
However, if time-reversal is not a symmetry of the microscopic system, such as in the presence of an external magnetic field $B$ breaking the symmetry, the Hamiltonian is subject to $\hat\Theta\hat H(B)\hat\Theta^{-1}=\hat H(-B)$. Consequently, the aforementioned equation transforms to
\begin{equation}\label{eqn:onsager_relations_general}
	G^R_{ab}(\omega,\vect{k};B)=\eta_a\eta_b G^R_{ba}(\omega,-\vect{k};-B) \,.
\end{equation}
This equation lies at the heart of the Onsager reciprocal relations \cite{Onsager:ReciprocalRelationsIrreversible,Onsager:ReciprocalRelationsIrreversiblea}, although older works approach it through thermodynamics and the relationships between fluxes and forces.

The requirement imposed by time-reversal on $G^R$ isn't automatically fulfilled by linearized hydrodynamics. Instead, it should be depicted as a constraint in the constitutive relations. When $\vect{k}\rightarrow0$ and $\omega=0$, this condition means that $\chi_T=0$ in first-order hydrodynamics \eqref{eqn:dissipative_current}. This serves as another method to constrain the transport coefficients without relying on the entropy current.

From \eqref{eqn:retarded_green_functions} and \eqref{eqn:onsager_relations_general}, it follows that the matrix $M_{ab}$ must adhere to
\begin{equation}
	\chi(B)SM^T(-\vect{k};-B)=M(\vect{k};B)\chi(B)S \,.
\end{equation}
In this equation, $S=\text{diag}\left(\eta_1,\eta_2,\dots\right)$ represents the matrix of time-reversal eigenvalues, and the susceptibilities adhere to the condition
\begin{equation}
	S\chi(B)S=\chi^T(-B) \,,
\end{equation}
One could apply a similar procedure to other discrete symmetries, such as parity. However, in these cases, the procedure is not as beneficial, as it is typically sufficient to monitor the parity-breaking parameters of the theory. This observation stems from the fact that only time-reversal symmetry is linked with an anti-unitary operator~\cite{Ammon:ChiralHydrodynamicsStrong}.
\subsection{Thermoelectric transport}\label{sec:thermoelectric_transport}
We aim to explore the thermoelectric transport coefficients within the framework of linear response theory. Typically, the charge and heat currents couple with the electric field and a temperature gradient, respectively. However, there are thermoelectric effects where the sources lead to mixed responses described by the complete thermoelectric matrix. In linear response, we have
\begin{equation}\label{eqn:thermoelectric_matrix}
	\begin{pmatrix}
		\delta J_i \\
		\delta Q_i
	\end{pmatrix}
	=
	\begin{pmatrix}
		\sigma_{ij} &   \alpha_{ij} \\
		T \bar{\alpha}_{ij} &   \bar{\kappa}_{ij}
	\end{pmatrix}
	\begin{pmatrix}
		\delta \mathbb{E}_j \\
		-\partial_j \delta T
	\end{pmatrix} \,,
\end{equation}
with $\sigma$ as the electrical conductivity tensor, $\bar{\kappa}$ as the thermal conductivity tensor, and $\alpha$ and $\bar{\alpha}$ as the thermoelectric tensors.

It's important to distinguish $\sigma$ from $\sigma_{ij}$: $\sigma$ is a transport coefficient determined by the microscopic theory, whereas $\sigma_{ij}$ characterizes the macroscopic response of the current to an external electric field and includes a contribution from $\sigma$.

It's noteworthy that $\bar{\kappa}$ differs from the conventional thermal conductivity measured in experiments. Laboratory measurements of thermal conductivity typically assume $\vect{J}=0$ as the boundary condition, whereas the above definition adopts $\vect{E}=0$ as the boundary condition. The relationship between $\kappa_{ij}$ and $\bar{\kappa}_{ij}$ is given by
\begin{equation}\label{eqn:thermal_conductivity_experiment}
\kappa_{ij}=\bar{\kappa}_{ij}-T\bar{\alpha}_{ik}\sigma_{kl}^{-1}\alpha_{lj} \,.
\end{equation}
Transport coefficients must adhere to Onsager relations, which implies that the matrices $\alpha_{ij}$ and $\bar{\alpha}_{ij}$ are connected as $\alpha_{ij}(B)=\bar{\alpha}_{ij}(B)$, where $B$ represents an external magnetic field that breaks time-reversal symmetry \cite{Hartnoll:TheoryNernstEffect}.

Our goal is to perturb the system with a small electric field and temperature gradient, and then determine how the electric and heat currents respond to these perturbations. To achieve this, we must first identify the appropriate sources. For instance, the electric current does not directly couple to the electric field in the Hamiltonian; instead, it couples to $A^{\mu}$.

A temperature perturbation, however, does couple to the Hamiltonian. Since we are operating within the grand canonical ensemble, we find
\begin{equation}
	\delta H=-\int\dif^d x\left(\frac{\delta T(t,\vect{x})}{T}(\epsilon(t,\vect{x})-\mu n(t,\vect{x}))+\delta A^{\mu}(t,\vect{x})J_{\mu}(t,\vect{x})\right) \,.
\end{equation}
We can set the gauge by defining the electric field as the negative gradient of $A^0$, $\mathbb{E}_i=-\partial_iA^0$, allowing us to neglect the $\delta A^i J_i$ terms and concentrate on $\delta A^0 n$
\begin{equation}
	\delta H=-\int\dif^d x\left(\frac{\delta T(t,\vect{x})}{T}(\epsilon(t,\vect{x})-\mu n(t,\vect{x}))+\delta A^{0}(t,\vect{x})n(t,\vect{x})\right) \,.
\end{equation}
 In classical thermodynamics, the heat current is typically defined as $\vect{J}_Q=T \vect{J}_S = \vect{J}_U - \mu \vect{J}_N$, where $\vect{J}_S$ represents the entropy current, while $\vect{J}_U$ and $\vect{J}_N$ denote the energy and charge currents, respectively. In the relativistic context, the heat current, in linear response theory, is
\begin{equation}\label{eqn_ch:2:canonical_heat_current}
	\delta Q^{i}=\delta T^{0 i}-\mu \delta J^{i} \,.
\end{equation}
Alternatively, the same result can be derived from a variational perspective by examining temperature fluctuations as alterations in the Euclidean time component of the metric. This approach, employed in \cite{Hartnoll:LecturesHolographicMethods,Herzog:LecturesHolographicSuperfluidity}, yields the canonical heat current form without relying on thermodynamics.

The Kubo formulae for the conductivities are expressed as
\begin{subequations}\label{eqn:thermoelectric_kubo_formulae}
	\begin{align}
		\sigma_{ij}(\omega)&=-\frac{G^R_{J_iJ_j}(\omega)-G^R_{J_iJ_j}(0)}{i\omega} \,,\\
		\alpha_{ij}(\omega)&=-\frac{G^R_{J_iQ_j}(\omega)-G^R_{J_iQ_j}(0)}{i\omega T} \,,\\
		\bar{\alpha}_{ij}(\omega)&=-\frac{G^R_{Q_iJ_j}(\omega)-G^R_{Q_iJ_j}(0)}{i\omega T} \,,\\
		\bar{\kappa}_{ij}(\omega)&=-\frac{G^R_{Q_iQ_j}(\omega)-G^R_{Q_iQ_j}(0)}{i\omega T} \,.
	\end{align}
\end{subequations}
From a practical standpoint, if our focus lies on the conductivities, we can simply introduce linear sources to perturb the system
\begin{equation}
	T\rightarrow T+\delta T-T x^i\delta\zeta_i \,,\qquad F^{0i}\rightarrow\delta\mathbb{E}^i \,.
\end{equation}
Here, $\zeta_i=\partial_iT/T$ represents the heat source. By solving the hydrodynamic equations in Laplace-Fourier space at $\vect{k}=0$, we can obtain \eqref{eqn:martin_kadanoff_kubo_formulae}, from which we can extract the conductivities \eqref{eqn:thermoelectric_kubo_formulae}.

Here are the optical conductivities in first-order hydrodynamics without a background magnetic field
\begin{subequations}\label{eqn:thermoelectric_conductivities}
\begin{align}
	\sigma(\omega)&=\sigma+\frac{in^2}{\omega(\epsilon+P)} \,,\\
	\alpha(\omega)&=-\frac{\mu}{T}\sigma+\frac{ins}{\omega(\epsilon+P)} \,,\\
	\bar\kappa(\omega)&=\frac{\mu^2}{T}\sigma+\frac{is^2T}{\omega(\epsilon+P)} \,.
\end{align}	
\end{subequations}
The Onsager relations impose the constraint $\bar\alpha=\alpha$. It's noteworthy that the conductivities diverge as $\omega$ approaches zero; notably, the imaginary part exhibits a pole. Therefore, according to Kramers-Kronig relations, the real part features a delta function at $\omega=0$ \cite{Hartnoll:HolographicQuantumMatter}. This divergence occurs because momentum serves as a conserved operator, implying that in the presence of a constant background electric field, a charged fluid accelerates infinitely. Lastly, the Green functions of conserved quantities adhere to specific Ward identities \cite{Herzog:LecturesHolographicSuperfluidity,Kovtun:LecturesHydrodynamicFluctuations}, which manifest in relations between the conductivities. In this scenario, it's straightforward to verify that
\begin{subequations}\label{eqn:ward_identities_conductivities}
\begin{align}
	\alpha(\omega)&=-\frac{\mu}{T}\sigma(\omega)+\frac{in}{T\omega} \,,\\
	\bar\kappa(\omega)&=\frac{\mu^2}{T}\sigma(\omega)+\frac{i\left(\epsilon+P-2\mu n\right)}{T\omega} \,.
\end{align}
\end{subequations}
Interestingly, in the presence of an external magnetic field $B$, these relations imply the existence of incoherent Hall conductivities \cite{Amoretti:2020mkp}.

\subsection{Modes and linear stability}
\label{sec:modes}
The first and most straightforward task in a hydrodynamic theory is to compute the modes. These modes provide insights into how small collective fluctuations near thermal equilibrium propagate within the fluid and offer crucial information about the theory's stability.

In a formal sense, the modes $\omega(\vect{k})$ represent the eigen-frequencies of the linearized equations of motion. Beginning with conservation laws and constitutive relations, we explore fluctuations around a global thermodynamic equilibrium state, characterized by
\begin{equation}
	\phi_a(t,\vect{x})=e^{i\omega t-i\vect{k}\cdot\vect{x}}\phi_a \,.
\end{equation}
The hydrodynamics equations can be expressed in Fourier space as
\begin{equation}
	M_{ab}(\vect{k})\phi_b=\omega\phi_a \,,
\end{equation}
which have non-trivial solutions only if $\text{det}(M-\mathds{1}\omega)=0$. From this determinant condition, we obtain the system's modes.

The causality requires that the imaginary part of the modes must be negative. This relates to the analyticity of the retarded correlators $G^R$ discussed in Section~\ref{sec:martin_kadanoff}, as the poles of the Green functions correspond to the modes themselves. This can be seen by noting that obtaining the Green functions involves inverting the hydrodynamic matrix, resulting in the determinant of $M-\mathds{1}\omega$ in the denominator, with the poles residing in the lower half of the complex $\omega$ plane.

Hydrodynamic modes are expected to satisfy $\omega(\vect{k}=0)=0$; however, in the presence of impurities, magnetic fields, or other slow variables influencing the hydrodynamic regime, this condition may not hold true \cite{Grozdanov:HolographyHydrodynamicsWeakly}. Therefore, it is customary to classify a mode as \emph{hydrodynamic} based on being the lowest-lying mode (closest to the real axis at $\vect{k}=0$), regardless of its specific value, as long as non-hydrodynamic modes are sufficiently deep in the lower half of the complex plane.
\section{Summary}
\label{sec:conclude}
In this review we have presented various approaches to relativistic hydrodynamics, which finds several applications in high-energy physics, condensed matter and astrophysics. Particularly, we have analysed in depth the canonical approach, focusing on the symmetries, thermodynamics and physical content. Finally, we have discussed thoroughly what happens when we restrict the dynamics to linearized perturbations, explaining how to obtain analytical expressions for the retarded correlators and thermoelectric transport using linear response theory.

Relaxed hydrodynamics, or quasihydrodynamics, acknowledges that real systems are rarely perfectly closed and free of impurities. Consequently, the hydrodynamic conservation laws are only approximate, typically broken by effective relaxation terms described by $\Gamma \sim 1/\tau_1$, dictating the damping of perturbations towards the state of equilibrium \cite{Martinoia:2024cbw}.

For hydrodynamics to remain valid, these corrections to the equations of conservation laws must be sufficiently small in magnitude, implying that whatever is disrupting exact conservation operates on scales much larger than the microscopic mean-free-path and the scale of thermalization (non-hydrodynamic modes). This allows for a separation of scales to persist.

In nuclear physics or astrophysics, systems like the quark-gluon-plasma formed in heavy-ion collisions may approximate perfect fluids with few impurities, limiting the applicability of quasihydrodynamics. However, condensed matter systems are more complex, with electrons interacting with a background lattice and phonons, and experiencing impurities and Umklapp scatterings that can disrupt their flow. Additionally, boundaries and layers contribute to losses of conserved charges, and approximate symmetries in condensed matter systems lead to explicitly weakly broken conservation equations~\cite{Amoretti:2023hpb}.

There are two main approaches to handle these corrections: 
\begin{enumerate}
    \item Including a Dynamical Mode: This approach assumes an extra mode that is very light and dynamical, quickly thermalizing. This extends the regime of validity of hydrodynamics to include this mode.
    \item Integrating Out the Mode: Here, the related microscopic non-hydrodynamical mode is integrated out, and its effect is represented by modifying the conservation law with a relaxation term.
\end{enumerate}
Another viewpoint, as proposed in \cite{Grozdanov:HolographyHydrodynamicsWeakly}, is that hydrodynamics operates on scales which are significantly larger than the microscopic mean free path $l$ and microscopic time $\tau$. While non-hydrodynamic operators typically decay quickly, there may be some operator with a parametrically slower decay rate. This allows for the inclusion of this operator as a quasi-conserved quantity with an effective decay time $\tau_1$, valid as long as $\tau_2 \partial_t \ll 1$. 

In summary, relaxed hydrodynamics recognizes the limits of perfect conservation in real systems and seeks to accommodate these deviations from exact conservation within the framework of hydrodynamics.

In this paper we concerned ourselves only with standard hydrodynamics, however there are many possible extensions that expand its regime of applicability to various different situations (e.g., dynamical magnetic fields, superfluid phases, quantum anomalies, spin polarization, etc.). Of particular interest are quasihydrodynamics~\cite{Martinoia:2024cbw,Amoretti:2023vhe} and spin magneto-hydrodynamics~\cite{Singh:2021man,Singh:2022ltu}.
\begin{acknowledgments}
We acknowledge Chirality 2024 conference in Timișoara, Romania where we started discussing about this note.
L.M. acknowledges support from
the project PRIN 2022ZTPK4E by the Italian Ministry of University and Research (MUR).
R.S. acknowledges kind hospitality and support of INFN Firenze, and Department of Physics and Astronomy, University of Florence where part of this work is completed.
R.S. is partly supported by the Florence University fellowship ``Effetti quantistici nei fluidi relativistici'' and by a postdoctoral fellowship of West University of Timișoara, Romania.
\end{acknowledgments}
\bibliography{pv_ref}{}

\providecommand{\href}[2]{#2}\begingroup\raggedright\begin{thebibliography}{100}

\bibitem{Landau:FluidMechanicsVolume}
L.~D. Landau and E.~M. Lifshitz, {\em Fluid {{Mechanics}}: {{Volume}} 6}.
\newblock Landau and {{Lifshitz}}: {{Course}} of {{Theoretical Physics}}. {Elsevier}, 1987.

\bibitem{Shibata:GeneralRelativisticViscous}
M.~Shibata, K.~Kiuchi, and Y.-i. Sekiguchi, ``{General relativistic viscous hydrodynamics of differentially rotating neutron stars},'' \href{http://dx.doi.org/10.1103/PhysRevD.95.083005}{{\em Phys. Rev. D} {\bfseries 95} no.~8, (2017) 083005}, \href{http://arxiv.org/abs/1703.10303}{{\ttfamily arXiv:1703.10303 [astro-ph.HE]}}.

\bibitem{Faber:HydrodynamicsNeutronStar}
J.~A. Faber and F.~A. Rasio, \href{http://dx.doi.org/10.1063/1.1387306}{``{Hydrodynamics of neutron star mergers},''} in {\em Astrophysical Sources for Ground-Based Gravitational Wave Detectors}, J.~M. Centrella, ed., vol.~575, pp.~130--142.
\newblock 2001.
\newblock \href{http://arxiv.org/abs/gr-qc/0101074}{{\ttfamily arXiv:gr-qc/0101074}}.

\bibitem{Balbus:InstabilityTurbulenceEnhanced}
S.~A. Balbus and J.~F. Hawley, ``{Instability, turbulence, and enhanced transport in accretion disks},'' \href{http://dx.doi.org/10.1103/RevModPhys.70.1}{{\em Rev. Mod. Phys.} {\bfseries 70} (1998) 1--53}.

\bibitem{Lucas:HydrodynamicsElectronsGraphene}
A.~Lucas and K.~C. Fong, ``{Hydrodynamics of electrons in graphene},'' \href{http://dx.doi.org/10.1088/1361-648X/aaa274}{{\em J. Phys. Condens. Matter} {\bfseries 30} no.~5, (2018) 053001}, \href{http://arxiv.org/abs/1710.08425}{{\ttfamily arXiv:1710.08425 [cond-mat.str-el]}}.

\bibitem{Narozhny:ElectronicHydrodynamicsGraphene}
B.~N. Narozhny, ``Electronic hydrodynamics in graphene,'' \href{http://dx.doi.org/10.1016/j.aop.2019.167979}{{\em Annals of Physics} {\bfseries 411} (Dec., 2019) 167979}.

\bibitem{Hartnoll:TheoryNernstEffect}
S.~A. Hartnoll, P.~K. Kovtun, M.~Muller, and S.~Sachdev, ``{Theory of the Nernst effect near quantum phase transitions in condensed matter, and in dyonic black holes},'' \href{http://dx.doi.org/10.1103/PhysRevB.76.144502}{{\em Phys. Rev. B} {\bfseries 76} (2007) 144502}, \href{http://arxiv.org/abs/0706.3215}{{\ttfamily arXiv:0706.3215 [cond-mat.str-el]}}.

\bibitem{Baggioli:ColloquiumHydrodynamicsHolography}
M.~Baggioli and B.~Gout\'eraux, ``{Colloquium: Hydrodynamics and holography of charge density wave phases},'' \href{http://dx.doi.org/10.1103/RevModPhys.95.011001}{{\em Rev. Mod. Phys.} {\bfseries 95} no.~1, (2023) 011001}, \href{http://arxiv.org/abs/2203.03298}{{\ttfamily arXiv:2203.03298 [hep-th]}}.

\bibitem{Lucas:HydrodynamicTheoryThermoelectric}
A.~Lucas, R.~A. Davison, and S.~Sachdev, ``{Hydrodynamic theory of thermoelectric transport and negative magnetoresistance in Weyl semimetals},'' \href{http://dx.doi.org/10.1073/pnas.1608881113}{{\em Proc. Nat. Acad. Sci.} {\bfseries 113} (2016) 9463}, \href{http://arxiv.org/abs/1604.08598}{{\ttfamily arXiv:1604.08598 [cond-mat.str-el]}}.

\bibitem{Amoretti:2019buu}
A.~Amoretti, M.~Meinero, D.~K. Brattan, F.~Caglieris, E.~Giannini, M.~Affronte, C.~Hess, B.~Buechner, N.~Magnoli, and M.~Putti, ``{Hydrodynamical description for magneto-transport in the strange metal phase of Bi-2201},'' \href{http://dx.doi.org/10.1103/PhysRevResearch.2.023387}{{\em Phys. Rev. Res.} {\bfseries 2} no.~2, (2020) 023387}, \href{http://arxiv.org/abs/1909.07991}{{\ttfamily arXiv:1909.07991 [cond-mat.str-el]}}.

\bibitem{Toner:FlocksHerdsSchools}
J.~Toner and Y.~Tu, ``Flocks, herds, and schools: {{A}} quantitative theory of flocking,'' \href{http://arxiv.org/abs/cond-mat/9804180}{{\ttfamily cond-mat/9804180}}.

\bibitem{Armas:2024iuy}
J.~Armas, A.~Jain, and R.~Lier, ``{Hydrodynamics of thermal active matter},'' \href{http://arxiv.org/abs/2405.11023}{{\ttfamily arXiv:2405.11023 [cond-mat.soft]}}.

\bibitem{Amoretti:2024obt}
A.~Amoretti, D.~K. Brattan, and L.~Martinoia, ``{Thermodynamic constraints and exact scaling exponents of flocking matter},'' \href{http://dx.doi.org/10.1103/PhysRevE.110.054108}{{\em Phys. Rev. E} {\bfseries 110} no.~5, (2024) 054108}, \href{http://arxiv.org/abs/2405.02283}{{\ttfamily arXiv:2405.02283 [cond-mat.stat-mech]}}.

\bibitem{Teaney:2009qa}
D.~A. Teaney, {\em {Viscous Hydrodynamics and the Quark Gluon Plasma}}, \href{http://dx.doi.org/10.1142/9789814293297_0004}{pp.~207--266}.
\newblock World Scientific, 2010.
\newblock \href{http://arxiv.org/abs/0905.2433}{{\ttfamily arXiv:0905.2433 [nucl-th]}}.

\bibitem{Palni:2024wdy}
P.~Palni {\em et~al.}, ``{Dynamics of Hot QCD Matter 2024 -- Bulk Properties},'' in {\em {2nd Hot QCD Matter Conference 2024}}.
\newblock 12, 2024.
\newblock \href{http://arxiv.org/abs/2412.10779}{{\ttfamily arXiv:2412.10779 [nucl-th]}}.

\bibitem{Arslandok:HotQCDWhite}
M.~Arslandok {\em et~al.}, ``{Hot QCD White Paper},'' \href{http://arxiv.org/abs/2303.17254}{{\ttfamily arXiv:2303.17254 [nucl-ex]}}.

\bibitem{Becattini:2024uha}
F.~Becattini, M.~Buzzegoli, T.~Niida, S.~Pu, A.-H. Tang, and Q.~Wang, ``{Spin polarization in relativistic heavy-ion collisions},'' \href{http://arxiv.org/abs/2402.04540}{{\ttfamily arXiv:2402.04540 [nucl-th]}}.

\bibitem{Heinz:CollectiveFlowViscosity}
U.~Heinz and R.~Snellings, ``{Collective flow and viscosity in relativistic heavy-ion collisions},'' \href{http://dx.doi.org/10.1146/annurev-nucl-102212-170540}{{\em Ann. Rev. Nucl. Part. Sci.} {\bfseries 63} (2013) 123--151}, \href{http://arxiv.org/abs/1301.2826}{{\ttfamily arXiv:1301.2826 [nucl-th]}}.

\bibitem{Rocha:2023ilf}
G.~S. Rocha, D.~Wagner, G.~S. Denicol, J.~Noronha, and D.~H. Rischke, ``{Theories of Relativistic Dissipative Fluid Dynamics},'' \href{http://dx.doi.org/10.3390/e26030189}{{\em Entropy} {\bfseries 26} no.~3, (2024) 189}, \href{http://arxiv.org/abs/2311.15063}{{\ttfamily arXiv:2311.15063 [nucl-th]}}.

\bibitem{Sorensen:2023zkk}
A.~Sorensen {\em et~al.}, ``{Dense nuclear matter equation of state from heavy-ion collisions},'' \href{http://dx.doi.org/10.1016/j.ppnp.2023.104080}{{\em Prog. Part. Nucl. Phys.} {\bfseries 134} (2024) 104080}, \href{http://arxiv.org/abs/2301.13253}{{\ttfamily arXiv:2301.13253 [nucl-th]}}.

\bibitem{Rangamani:GravityHydrodynamicsLectures}
M.~Rangamani, ``{Gravity and Hydrodynamics: Lectures on the fluid-gravity correspondence},'' \href{http://dx.doi.org/10.1088/0264-9381/26/22/224003}{{\em Class. Quant. Grav.} {\bfseries 26} (2009) 224003}, \href{http://arxiv.org/abs/0905.4352}{{\ttfamily arXiv:0905.4352 [hep-th]}}.

\bibitem{Hartnoll:HolographicQuantumMatter}
S.~A. Hartnoll, A.~Lucas, and S.~Sachdev, {\em {Holographic quantum matter}}.
\newblock The MIT Press, 12, 2016.
\newblock \href{http://arxiv.org/abs/1612.07324}{{\ttfamily arXiv:1612.07324 [hep-th]}}.

\bibitem{Bhattacharya:TheoryFirstOrder}
J.~Bhattacharya, S.~Bhattacharyya, S.~Minwalla, and A.~Yarom, ``{A Theory of first order dissipative superfluid dynamics},'' \href{http://dx.doi.org/10.1007/JHEP05(2014)147}{{\em JHEP} {\bfseries 05} (2014) 147}, \href{http://arxiv.org/abs/1105.3733}{{\ttfamily arXiv:1105.3733 [hep-th]}}.

\bibitem{Baier:RelativisticViscousHydrodynamics}
R.~Baier, P.~Romatschke, D.~T. Son, A.~O. Starinets, and M.~A. Stephanov, ``{Relativistic viscous hydrodynamics, conformal invariance, and holography},'' \href{http://dx.doi.org/10.1088/1126-6708/2008/04/100}{{\em JHEP} {\bfseries 04} (2008) 100}, \href{http://arxiv.org/abs/0712.2451}{{\ttfamily arXiv:0712.2451 [hep-th]}}.

\bibitem{Bhattacharyya:NonlinearFluidDynamics}
S.~Bhattacharyya, V.~E. Hubeny, S.~Minwalla, and M.~Rangamani, ``{Nonlinear Fluid Dynamics from Gravity},'' \href{http://dx.doi.org/10.1088/1126-6708/2008/02/045}{{\em JHEP} {\bfseries 02} (2008) 045}, \href{http://arxiv.org/abs/0712.2456}{{\ttfamily arXiv:0712.2456 [hep-th]}}.

\bibitem{Romatschke:RelativisticViscousFluid}
P.~Romatschke, ``{Relativistic Viscous Fluid Dynamics and Non-Equilibrium Entropy},'' \href{http://dx.doi.org/10.1088/0264-9381/27/2/025006}{{\em Class. Quant. Grav.} {\bfseries 27} (2010) 025006}, \href{http://arxiv.org/abs/0906.4787}{{\ttfamily arXiv:0906.4787 [hep-th]}}.

\bibitem{El:ThirdorderRelativisticDissipative}
A.~El, Z.~Xu, and C.~Greiner, ``{Third-order relativistic dissipative hydrodynamics},'' \href{http://dx.doi.org/10.1103/PhysRevC.81.041901}{{\em Phys. Rev. C} {\bfseries 81} (2010) 041901}, \href{http://arxiv.org/abs/0907.4500}{{\ttfamily arXiv:0907.4500 [hep-ph]}}.

\bibitem{Grozdanov:ConstructingHigherorderHydrodynamics}
S.~Grozdanov and N.~Kaplis, ``{Constructing higher-order hydrodynamics: The third order},'' \href{http://dx.doi.org/10.1103/PhysRevD.93.066012}{{\em Phys. Rev. D} {\bfseries 93} no.~6, (2016) 066012}, \href{http://arxiv.org/abs/1507.02461}{{\ttfamily arXiv:1507.02461 [hep-th]}}.

\bibitem{Jensen:ParityViolatingHydrodynamicsDimensions}
K.~Jensen, M.~Kaminski, P.~Kovtun, R.~Meyer, A.~Ritz, and A.~Yarom, ``{Parity-Violating Hydrodynamics in 2+1 Dimensions},'' \href{http://dx.doi.org/10.1007/JHEP05(2012)102}{{\em JHEP} {\bfseries 05} (2012) 102}, \href{http://arxiv.org/abs/1112.4498}{{\ttfamily arXiv:1112.4498 [hep-th]}}.

\bibitem{Lier:PassiveOddViscoelasticity}
R.~Lier, J.~Armas, S.~Bo, C.~Duclut, F.~J\"ulicher, and P.~Sur\'owka, ``{Passive odd viscoelasticity},'' \href{http://dx.doi.org/10.1103/PhysRevE.105.054607}{{\em Phys. Rev. E} {\bfseries 105} no.~5, (2022) 054607}, \href{http://arxiv.org/abs/2109.06606}{{\ttfamily arXiv:2109.06606 [cond-mat.soft]}}.

\bibitem{Lucas:PhenomenologyNonrelativisticParityviolating}
A.~Lucas and P.~Sur\'owka, ``{Phenomenology of nonrelativistic parity-violating hydrodynamics in 2+1 dimensions},'' \href{http://dx.doi.org/10.1103/PhysRevE.90.063005}{{\em Phys. Rev. E} {\bfseries 90} no.~6, (2014) 063005}, \href{http://arxiv.org/abs/1403.5239}{{\ttfamily arXiv:1403.5239 [cond-mat.mes-hall]}}.

\bibitem{Son:HydrodynamicsTriangleAnomalies}
D.~T. Son and P.~Surowka, ``{Hydrodynamics with Triangle Anomalies},'' \href{http://dx.doi.org/10.1103/PhysRevLett.103.191601}{{\em Phys. Rev. Lett.} {\bfseries 103} (2009) 191601}, \href{http://arxiv.org/abs/0906.5044}{{\ttfamily arXiv:0906.5044 [hep-th]}}.

\bibitem{Jensen:TriangleAnomaliesThermodynamics}
K.~Jensen, ``{Triangle Anomalies, Thermodynamics, and Hydrodynamics},'' \href{http://dx.doi.org/10.1103/PhysRevD.85.125017}{{\em Phys. Rev. D} {\bfseries 85} (2012) 125017}, \href{http://arxiv.org/abs/1203.3599}{{\ttfamily arXiv:1203.3599 [hep-th]}}.

\bibitem{Amoretti:2022vxq}
A.~Amoretti, D.~K. Brattan, L.~Martinoia, and I.~Matthaiakakis, ``{Leading order magnetic field dependence of conductivities in anomalous hydrodynamics},'' \href{http://dx.doi.org/10.1103/PhysRevD.108.016003}{{\em Phys. Rev. D} {\bfseries 108} no.~1, (2023) 016003}, \href{http://arxiv.org/abs/2212.09761}{{\ttfamily arXiv:2212.09761 [hep-th]}}.

\bibitem{Gallegos:HydrodynamicsSpinCurrents}
A.~D. Gallegos, U.~G\"ursoy, and A.~Yarom, ``{Hydrodynamics of spin currents},'' \href{http://dx.doi.org/10.21468/SciPostPhys.11.2.041}{{\em SciPost Phys.} {\bfseries 11} (2021) 041}, \href{http://arxiv.org/abs/2101.04759}{{\ttfamily arXiv:2101.04759 [hep-th]}}.

\bibitem{Becattini:SpinTensorIts}
F.~Becattini, W.~Florkowski, and E.~Speranza, ``{Spin tensor and its role in non-equilibrium thermodynamics},'' \href{http://dx.doi.org/10.1016/j.physletb.2018.12.016}{{\em Phys. Lett. B} {\bfseries 789} (2019) 419--425}, \href{http://arxiv.org/abs/1807.10994}{{\ttfamily arXiv:1807.10994 [hep-th]}}.

\bibitem{Florkowski:2018fap}
W.~Florkowski, A.~Kumar, and R.~Ryblewski, ``{Relativistic hydrodynamics for spin-polarized fluids},'' \href{http://dx.doi.org/10.1016/j.ppnp.2019.07.001}{{\em Prog. Part. Nucl. Phys.} {\bfseries 108} (2019) 103709}, \href{http://arxiv.org/abs/1811.04409}{{\ttfamily arXiv:1811.04409 [nucl-th]}}.

\bibitem{Florkowski:2019qdp}
W.~Florkowski, A.~Kumar, R.~Ryblewski, and R.~Singh, ``{Spin polarization evolution in a boost invariant hydrodynamical background},'' \href{http://dx.doi.org/10.1103/PhysRevC.99.044910}{{\em Phys. Rev. C} {\bfseries 99} no.~4, (2019) 044910}, \href{http://arxiv.org/abs/1901.09655}{{\ttfamily arXiv:1901.09655 [hep-ph]}}.

\bibitem{Florkowski:2021wvk}
W.~Florkowski, R.~Ryblewski, R.~Singh, and G.~Sophys, ``{Spin polarization dynamics in the non-boost-invariant background},'' \href{http://dx.doi.org/10.1103/PhysRevD.105.054007}{{\em Phys. Rev. D} {\bfseries 105} no.~5, (2022) 054007}, \href{http://arxiv.org/abs/2112.01856}{{\ttfamily arXiv:2112.01856 [hep-ph]}}.

\bibitem{Singh:2020rht}
R.~Singh, G.~Sophys, and R.~Ryblewski, ``{Spin polarization dynamics in the Gubser-expanding background},'' \href{http://dx.doi.org/10.1103/PhysRevD.103.074024}{{\em Phys. Rev. D} {\bfseries 103} no.~7, (2021) 074024}, \href{http://arxiv.org/abs/2011.14907}{{\ttfamily arXiv:2011.14907 [hep-ph]}}.

\bibitem{Singh:2022uyy}
R.~Singh, ``{Collective dynamics of polarized spin-half fermions in relativistic heavy-ion collisions},'' \href{http://dx.doi.org/10.1142/S0217751X23300119}{{\em Int. J. Mod. Phys. A} {\bfseries 38} no.~20, (2023) 2330011}, \href{http://arxiv.org/abs/2212.06569}{{\ttfamily arXiv:2212.06569 [hep-ph]}}.

\bibitem{Speranza:2020ilk}
E.~Speranza and N.~Weickgenannt, ``{Spin tensor and pseudo-gauges: from nuclear collisions to gravitational physics},'' \href{http://dx.doi.org/10.1140/epja/s10050-021-00455-2}{{\em Eur. Phys. J. A} {\bfseries 57} no.~5, (2021) 155}, \href{http://arxiv.org/abs/2007.00138}{{\ttfamily arXiv:2007.00138 [nucl-th]}}.

\bibitem{Bhadury:2020puc}
S.~Bhadury, W.~Florkowski, A.~Jaiswal, A.~Kumar, and R.~Ryblewski, ``{Relativistic dissipative spin dynamics in the relaxation time approximation},'' \href{http://dx.doi.org/10.1016/j.physletb.2021.136096}{{\em Phys. Lett. B} {\bfseries 814} (2021) 136096}, \href{http://arxiv.org/abs/2002.03937}{{\ttfamily arXiv:2002.03937 [hep-ph]}}.

\bibitem{Bhadury:2021oat}
S.~Bhadury, J.~Bhatt, A.~Jaiswal, and A.~Kumar, ``{New developments in relativistic fluid dynamics with spin},'' \href{http://dx.doi.org/10.1140/epjs/s11734-021-00020-4}{{\em Eur. Phys. J. ST} {\bfseries 230} no.~3, (2021) 655--672}, \href{http://arxiv.org/abs/2101.11964}{{\ttfamily arXiv:2101.11964 [hep-ph]}}.

\bibitem{Bhadury:2022ulr}
S.~Bhadury, W.~Florkowski, A.~Jaiswal, A.~Kumar, and R.~Ryblewski, ``{Relativistic Spin Magnetohydrodynamics},'' \href{http://dx.doi.org/10.1103/PhysRevLett.129.192301}{{\em Phys. Rev. Lett.} {\bfseries 129} no.~19, (2022) 192301}, \href{http://arxiv.org/abs/2204.01357}{{\ttfamily arXiv:2204.01357 [nucl-th]}}.

\bibitem{Amoretti:2017axe}
A.~Amoretti, D.~Are\'an, B.~Gout\'eraux, and D.~Musso, ``{DC resistivity of quantum critical, charge density wave states from gauge-gravity duality},'' \href{http://dx.doi.org/10.1103/PhysRevLett.120.171603}{{\em Phys. Rev. Lett.} {\bfseries 120} no.~17, (2018) 171603}, \href{http://arxiv.org/abs/1712.07994}{{\ttfamily arXiv:1712.07994 [hep-th]}}.

\bibitem{Amoretti:2017frz}
A.~Amoretti, D.~Are\'an, B.~Gout\'eraux, and D.~Musso, ``{Effective holographic theory of charge density waves},'' \href{http://dx.doi.org/10.1103/PhysRevD.97.086017}{{\em Phys. Rev. D} {\bfseries 97} no.~8, (2018) 086017}, \href{http://arxiv.org/abs/1711.06610}{{\ttfamily arXiv:1711.06610 [hep-th]}}.

\bibitem{Amoretti:2018tzw}
A.~Amoretti, D.~Are\'an, B.~Gout\'eraux, and D.~Musso, ``{Universal relaxation in a holographic metallic density wave phase},'' \href{http://dx.doi.org/10.1103/PhysRevLett.123.211602}{{\em Phys. Rev. Lett.} {\bfseries 123} no.~21, (2019) 211602}, \href{http://arxiv.org/abs/1812.08118}{{\ttfamily arXiv:1812.08118 [hep-th]}}.

\bibitem{Armas:ApproximateSymmetriesPseudoGoldstones}
J.~Armas, A.~Jain, and R.~Lier, ``{Approximate symmetries, pseudo-Goldstones, and the second law of thermodynamics},'' \href{http://dx.doi.org/10.1103/PhysRevD.108.086011}{{\em Phys. Rev. D} {\bfseries 108} no.~8, (2023) 086011}, \href{http://arxiv.org/abs/2112.14373}{{\ttfamily arXiv:2112.14373 [hep-th]}}.

\bibitem{Armas:HydrodynamicsPlasticDeformations}
J.~Armas, E.~van Heumen, A.~Jain, and R.~Lier, ``{Hydrodynamics of plastic deformations in electronic crystals},'' \href{http://dx.doi.org/10.1103/PhysRevB.107.155108}{{\em Phys. Rev. B} {\bfseries 107} no.~15, (2023) 155108}, \href{http://arxiv.org/abs/2211.02117}{{\ttfamily arXiv:2211.02117 [cond-mat.str-el]}}.

\bibitem{Armas:HydrodynamicsChargeDensity}
J.~Armas and A.~Jain, ``{Hydrodynamics for charge density waves and their holographic duals},'' \href{http://dx.doi.org/10.1103/PhysRevD.101.121901}{{\em Phys. Rev. D} {\bfseries 101} no.~12, (2020) 121901}, \href{http://arxiv.org/abs/2001.07357}{{\ttfamily arXiv:2001.07357 [hep-th]}}.

\bibitem{Amoretti:2021lll}
A.~Amoretti, D.~Arean, D.~K. Brattan, and L.~Martinoia, ``{Hydrodynamic magneto-transport in holographic charge density wave states},'' \href{http://dx.doi.org/10.1007/JHEP11(2021)011}{{\em JHEP} {\bfseries 11} (2021) 011}, \href{http://arxiv.org/abs/2107.00519}{{\ttfamily arXiv:2107.00519 [hep-th]}}.

\bibitem{Amoretti:2021fch}
A.~Amoretti, D.~Arean, D.~K. Brattan, and N.~Magnoli, ``{Hydrodynamic magneto-transport in charge density wave states},'' \href{http://dx.doi.org/10.1007/JHEP05(2021)027}{{\em JHEP} {\bfseries 05} (2021) 027}, \href{http://arxiv.org/abs/2101.05343}{{\ttfamily arXiv:2101.05343 [hep-th]}}.

\bibitem{Amoretti:2022acb}
A.~Amoretti and D.~K. Brattan, ``{On the hydrodynamics of (2 + 1)-dimensional strongly coupled relativistic theories in an external magnetic field},'' \href{http://dx.doi.org/10.1142/S0217732322300105}{{\em Mod. Phys. Lett. A} {\bfseries 37} no.~21, (2022) 2230010}, \href{http://arxiv.org/abs/2209.11589}{{\ttfamily arXiv:2209.11589 [hep-th]}}.

\bibitem{Kovtun:EffectiveActionRelativistic}
P.~Kovtun, G.~D. Moore, and P.~Romatschke, ``{Towards an effective action for relativistic dissipative hydrodynamics},'' \href{http://dx.doi.org/10.1007/JHEP07(2014)123}{{\em JHEP} {\bfseries 07} (2014) 123}, \href{http://arxiv.org/abs/1405.3967}{{\ttfamily arXiv:1405.3967 [hep-ph]}}.

\bibitem{Grozdanov:ViscosityDissipativeHydrodynamics}
S.~Grozdanov and J.~Polonyi, ``{Viscosity and dissipative hydrodynamics from effective field theory},'' \href{http://dx.doi.org/10.1103/PhysRevD.91.105031}{{\em Phys. Rev. D} {\bfseries 91} no.~10, (2015) 105031}, \href{http://arxiv.org/abs/1305.3670}{{\ttfamily arXiv:1305.3670 [hep-th]}}.

\bibitem{Glorioso:LecturesNonequilibriumEffective}
H.~Liu and P.~Glorioso, ``{Lectures on non-equilibrium effective field theories and fluctuating hydrodynamics},'' \href{http://dx.doi.org/10.22323/1.305.0008}{{\em PoS} {\bfseries TASI2017} (2018) 008}, \href{http://arxiv.org/abs/1805.09331}{{\ttfamily arXiv:1805.09331 [hep-th]}}.

\bibitem{Haehl:FluidManifestoEmergent}
F.~M. Haehl, R.~Loganayagam, P.~Narayan, A.~A. Nizami, and M.~Rangamani, ``{Thermal out-of-time-order correlators, KMS relations, and spectral functions},'' \href{http://dx.doi.org/10.1007/JHEP12(2017)154}{{\em JHEP} {\bfseries 12} (2017) 154}, \href{http://arxiv.org/abs/1706.08956}{{\ttfamily arXiv:1706.08956 [hep-th]}}.

\bibitem{Haehl:EffectiveActionRelativistic}
F.~M. Haehl, R.~Loganayagam, and M.~Rangamani, ``{Effective Action for Relativistic Hydrodynamics: Fluctuations, Dissipation, and Entropy Inflow},'' \href{http://dx.doi.org/10.1007/JHEP10(2018)194}{{\em JHEP} {\bfseries 10} (2018) 194}, \href{http://arxiv.org/abs/1803.11155}{{\ttfamily arXiv:1803.11155 [hep-th]}}.

\bibitem{Haehl:AdiabaticHydrodynamicsEightfold}
F.~M. Haehl, R.~Loganayagam, and M.~Rangamani, ``{Adiabatic hydrodynamics: The eightfold way to dissipation},'' \href{http://dx.doi.org/10.1007/JHEP05(2015)060}{{\em JHEP} {\bfseries 05} (2015) 060}, \href{http://arxiv.org/abs/1502.00636}{{\ttfamily arXiv:1502.00636 [hep-th]}}.

\bibitem{Haehl:EightfoldWayDissipation}
F.~M. Haehl, R.~Loganayagam, and M.~Rangamani, ``{The eightfold way to dissipation},'' \href{http://dx.doi.org/10.1103/PhysRevLett.114.201601}{{\em Phys. Rev. Lett.} {\bfseries 114} (2015) 201601}, \href{http://arxiv.org/abs/1412.1090}{{\ttfamily arXiv:1412.1090 [hep-th]}}.

\bibitem{Baggioli:QuasihydrodynamicsSchwingerKeldyshEffective}
M.~Baggioli, Y.~Bu, and V.~Ziogas, ``{U(1) quasi-hydrodynamics: Schwinger-Keldysh effective field theory and holography},'' \href{http://dx.doi.org/10.1007/JHEP09(2023)019}{{\em JHEP} {\bfseries 09} (2023) 019}, \href{http://arxiv.org/abs/2304.14173}{{\ttfamily arXiv:2304.14173 [hep-th]}}.

\bibitem{Grozdanov:HolographyHydrodynamicsWeakly}
S.~Grozdanov, A.~Lucas, and N.~Poovuttikul, ``{Holography and hydrodynamics with weakly broken symmetries},'' \href{http://dx.doi.org/10.1103/PhysRevD.99.086012}{{\em Phys. Rev. D} {\bfseries 99} no.~8, (2019) 086012}, \href{http://arxiv.org/abs/1810.10016}{{\ttfamily arXiv:1810.10016 [hep-th]}}.

\bibitem{Stephanov:HydroHydrodynamicsParametric}
M.~Stephanov and Y.~Yin, ``{Hydrodynamics with parametric slowing down and fluctuations near the critical point},'' \href{http://dx.doi.org/10.1103/PhysRevD.98.036006}{{\em Phys. Rev. D} {\bfseries 98} no.~3, (2018) 036006}, \href{http://arxiv.org/abs/1712.10305}{{\ttfamily arXiv:1712.10305 [nucl-th]}}.

\bibitem{Glodkowski:HydrodynamicsDipoleconservingFluids}
A.~G\l{}\'odkowski, F.~Pe\~na Ben\'\i{}tez, and P.~Sur\'owka, ``{Hydrodynamics of dipole-conserving fluids},'' \href{http://dx.doi.org/10.1103/PhysRevE.107.034142}{{\em Phys. Rev. E} {\bfseries 107} no.~3, (2023) 034142}, \href{http://arxiv.org/abs/2212.06848}{{\ttfamily arXiv:2212.06848 [cond-mat.str-el]}}.

\bibitem{Grosvenor:HydrodynamicsIdealFracton}
K.~T. Grosvenor, C.~Hoyos, F.~Pe\~na Ben\'\i{}tez, and P.~Sur\'owka, ``{Hydrodynamics of ideal fracton fluids},'' \href{http://dx.doi.org/10.1103/PhysRevResearch.3.043186}{{\em Phys. Rev. Res.} {\bfseries 3} no.~4, (2021) 043186}, \href{http://arxiv.org/abs/2105.01084}{{\ttfamily arXiv:2105.01084 [cond-mat.str-el]}}.

\bibitem{Guo:FractonHydrodynamicsTimereversal}
J.~Guo, P.~Glorioso, and A.~Lucas, ``{Fracton Hydrodynamics without Time-Reversal Symmetry},'' \href{http://dx.doi.org/10.1103/PhysRevLett.129.150603}{{\em Phys. Rev. Lett.} {\bfseries 129} no.~15, (2022) 150603}, \href{http://arxiv.org/abs/2204.06006}{{\ttfamily arXiv:2204.06006 [cond-mat.stat-mech]}}.

\bibitem{Das:HigherformSymmetriesAnomalous}
A.~Das, R.~Gregory, and N.~Iqbal, ``{Higher-form symmetries, anomalous magnetohydrodynamics, and holography},'' \href{http://dx.doi.org/10.21468/SciPostPhys.14.6.163}{{\em SciPost Phys.} {\bfseries 14} no.~6, (2023) 163}, \href{http://arxiv.org/abs/2205.03619}{{\ttfamily arXiv:2205.03619 [hep-th]}}.

\bibitem{Grozdanov:GeneralizedGlobalSymmetries}
S.~Grozdanov, D.~M. Hofman, and N.~Iqbal, ``{Generalized global symmetries and dissipative magnetohydrodynamics},'' \href{http://dx.doi.org/10.1103/PhysRevD.95.096003}{{\em Phys. Rev. D} {\bfseries 95} no.~9, (2017) 096003}, \href{http://arxiv.org/abs/1610.07392}{{\ttfamily arXiv:1610.07392 [hep-th]}}.

\bibitem{Armas:ApproximateHigherformSymmetries}
J.~Armas and A.~Jain, ``{Approximate higher-form symmetries, topological defects, and dynamical phase transitions},'' \href{http://dx.doi.org/10.1103/PhysRevD.109.045019}{{\em Phys. Rev. D} {\bfseries 109} no.~4, (2024) 045019}, \href{http://arxiv.org/abs/2301.09628}{{\ttfamily arXiv:2301.09628 [hep-th]}}.

\bibitem{Amoretti:2022ovc}
A.~Amoretti, D.~K. Brattan, L.~Martinoia, and I.~Matthaiakakis, ``{Non-dissipative electrically driven fluids},'' \href{http://dx.doi.org/10.1007/JHEP05(2023)218}{{\em JHEP} {\bfseries 05} (2023) 218}, \href{http://arxiv.org/abs/2211.05791}{{\ttfamily arXiv:2211.05791 [hep-th]}}.

\bibitem{Amoretti:2024jig}
A.~Amoretti, D.~K. Brattan, L.~Martinoia, and J.~Rongen, ``{Dissipative electrically driven fluids},'' \href{http://dx.doi.org/10.1007/JHEP12(2024)114}{{\em JHEP} {\bfseries 12} (2024) 114}, \href{http://arxiv.org/abs/2407.18856}{{\ttfamily arXiv:2407.18856 [cond-mat.stat-mech]}}.

\bibitem{Brattan:2024dfv}
D.~K. Brattan, M.~Matsumoto, M.~Baggioli, and A.~Amoretti, ``{Relaxed hydrodynamic theory of electrically driven non-equilibrium steady states},'' \href{http://dx.doi.org/10.1103/PhysRevResearch.6.043097}{{\em Phys. Rev. Res.} {\bfseries 6} (4, 2024) 043097}, \href{http://arxiv.org/abs/2404.05568}{{\ttfamily arXiv:2404.05568 [cond-mat.stat-mech]}}.

\bibitem{Jackiw:2004nm}
R.~Jackiw, V.~P. Nair, S.~Y. Pi, and A.~P. Polychronakos, ``{Perfect fluid theory and its extensions},'' \href{http://dx.doi.org/10.1088/0305-4470/37/42/R01}{{\em J. Phys. A} {\bfseries 37} (2004) R327--R432}, \href{http://arxiv.org/abs/hep-ph/0407101}{{\ttfamily arXiv:hep-ph/0407101}}.

\bibitem{Nair:2011mk}
V.~P. Nair, R.~Ray, and S.~Roy, ``{Fluids, Anomalies and the Chiral Magnetic Effect: A Group-Theoretic Formulation},'' \href{http://dx.doi.org/10.1103/PhysRevD.86.025012}{{\em Phys. Rev. D} {\bfseries 86} (2012) 025012}, \href{http://arxiv.org/abs/1112.4022}{{\ttfamily arXiv:1112.4022 [hep-th]}}.

\bibitem{Karabali:2014vla}
D.~Karabali and V.~P. Nair, ``{Relativistic Particle and Relativistic Fluids: Magnetic Moment and Spin-Orbit Interactions},'' \href{http://dx.doi.org/10.1103/PhysRevD.90.105018}{{\em Phys. Rev. D} {\bfseries 90} no.~10, (2014) 105018}, \href{http://arxiv.org/abs/1406.1551}{{\ttfamily arXiv:1406.1551 [hep-th]}}.

\bibitem{Monteiro:2014wsa}
G.~M. Monteiro, A.~G. Abanov, and V.~P. Nair, ``{Hydrodynamics with gauge anomaly: Variational principle and Hamiltonian formulation},'' \href{http://dx.doi.org/10.1103/PhysRevD.91.125033}{{\em Phys. Rev. D} {\bfseries 91} no.~12, (2015) 125033}, \href{http://arxiv.org/abs/1410.4833}{{\ttfamily arXiv:1410.4833 [hep-th]}}.

\bibitem{Nair:2020kjg}
V.~P. Nair, ``{Topological terms and diffeomorphism anomalies in fluid dynamics and sigma models},'' \href{http://dx.doi.org/10.1103/PhysRevD.103.085017}{{\em Phys. Rev. D} {\bfseries 103} no.~8, (2021) 085017}, \href{http://arxiv.org/abs/2008.11260}{{\ttfamily arXiv:2008.11260 [hep-th]}}.

\bibitem{Martinoia:2024cbw}
L.~Martinoia, \href{http://dx.doi.org/10.15167/martinoia-luca_phd2024-03-01}{{\em {Developments in quasihydrodynamics}}}.
\newblock PhD thesis, Genoa U., 2024.
\newblock \href{http://arxiv.org/abs/2403.14254}{{\ttfamily arXiv:2403.14254 [hep-th]}}.

\bibitem{Heller:HydrodynamicsGradientExpansion}
M.~P. Heller and M.~Spalinski, ``{Hydrodynamics Beyond the Gradient Expansion: Resurgence and Resummation},'' \href{http://dx.doi.org/10.1103/PhysRevLett.115.072501}{{\em Phys. Rev. Lett.} {\bfseries 115} no.~7, (2015) 072501}, \href{http://arxiv.org/abs/1503.07514}{{\ttfamily arXiv:1503.07514 [hep-th]}}.

\bibitem{Heller:2021oxl}
M.~P. Heller, A.~Serantes, M.~Spali\'nski, V.~Svensson, and B.~Withers, ``{Hydrodynamic Gradient Expansion Diverges beyond Bjorken Flow},'' \href{http://dx.doi.org/10.1103/PhysRevLett.128.122302}{{\em Phys. Rev. Lett.} {\bfseries 128} no.~12, (2022) 122302}, \href{http://arxiv.org/abs/2110.07621}{{\ttfamily arXiv:2110.07621 [hep-th]}}.

\bibitem{Hiscock:GenericInstabilitiesFirstorder}
W.~A. Hiscock and L.~Lindblom, ``{Generic instabilities in first-order dissipative relativistic fluid theories},'' \href{http://dx.doi.org/10.1103/PhysRevD.31.725}{{\em Phys. Rev. D} {\bfseries 31} (1985) 725--733}.

\bibitem{Hiscock:LinearPlaneWaves}
W.~A. Hiscock and L.~Lindblom, ``{Linear plane waves in dissipative relativistic fluids},'' \href{http://dx.doi.org/10.1103/PhysRevD.35.3723}{{\em Phys. Rev. D} {\bfseries 35} (1987) 3723--3732}.

\bibitem{Speranza:ChallengesSolvingChiral}
E.~Speranza, F.~S. Bemfica, M.~M. Disconzi, and J.~Noronha, ``{Challenges in solving chiral hydrodynamics},'' \href{http://dx.doi.org/10.1103/PhysRevD.107.054029}{{\em Phys. Rev. D} {\bfseries 107} no.~5, (2023) 054029}, \href{http://arxiv.org/abs/2104.02110}{{\ttfamily arXiv:2104.02110 [hep-th]}}.

\bibitem{Heller:2022ejw}
M.~P. Heller, A.~Serantes, M.~Spali\'nski, and B.~Withers, ``{Rigorous Bounds on Transport from Causality},'' \href{http://dx.doi.org/10.1103/PhysRevLett.130.261601}{{\em Phys. Rev. Lett.} {\bfseries 130} no.~26, (2023) 261601}, \href{http://arxiv.org/abs/2212.07434}{{\ttfamily arXiv:2212.07434 [hep-th]}}.

\bibitem{Heller:2021yjh}
M.~P. Heller, A.~Serantes, M.~Spali\'nski, V.~Svensson, and B.~Withers, ``{Relativistic Hydrodynamics: A Singulant Perspective},'' \href{http://dx.doi.org/10.1103/PhysRevX.12.041010}{{\em Phys. Rev. X} {\bfseries 12} no.~4, (2022) 041010}, \href{http://arxiv.org/abs/2112.12794}{{\ttfamily arXiv:2112.12794 [hep-th]}}.

\bibitem{Maxwell:IVDynamicalTheory}
J.~C. Maxwell, ``{{IV}}. {{On}} the dynamical theory of gases,''.

\bibitem{Cattaneo:SullaConduzioneCalore}
C.~Cattaneo, \href{http://dx.doi.org/10.1007/978-3-642-11051-1_5}{``Sulla {{Conduzione Del Calore}},''} in {\em Some {{Aspects}} of {{Diffusion Theory}}}, A.~Pignedoli, ed., C.{{I}}.{{M}}.{{E}}. {{Summer Schools}}.
\newblock {Springer}, 2011.

\bibitem{Mueller:ParadoxonWaermeleitungstheorie}
I.~M\"uller, ``Zum paradoxon der w\"armeleitungstheorie,''.

\bibitem{Israel:TransientRelativisticThermodynamics}
W.~Israel and J.~M. Stewart, ``Transient relativistic thermodynamics and kinetic theory,''.

\bibitem{Israel:NonstationaryIrreversibleThermodynamics}
W.~Israel, ``Nonstationary irreversible thermodynamics: {{A}} causal relativistic theory,''.

\bibitem{Romatschke:RelativisticFluidDynamics}
P.~Romatschke and U.~Romatschke, \href{http://dx.doi.org/10.1017/9781108651998}{{\em {Relativistic Fluid Dynamics In and Out of Equilibrium}}}.
\newblock Cambridge Monographs on Mathematical Physics. Cambridge University Press, 5, 2019.
\newblock \href{http://arxiv.org/abs/1712.05815}{{\ttfamily arXiv:1712.05815 [nucl-th]}}.

\bibitem{Bemfica:NonlinearCausalityGeneral}
F.~S. Bemfica, F.~S. Bemfica, M.~M. Disconzi, M.~M. Disconzi, J.~Noronha, and J.~Noronha, ``{Nonlinear Causality of General First-Order Relativistic Viscous Hydrodynamics},'' \href{http://dx.doi.org/10.1103/PhysRevD.100.104020}{{\em Phys. Rev. D} {\bfseries 100} no.~10, (2019) 104020}, \href{http://arxiv.org/abs/1907.12695}{{\ttfamily arXiv:1907.12695 [gr-qc]}}. [Erratum: Phys.Rev.D 105, 069902 (2022)].

\bibitem{Bemfica:CausalityExistenceSolutions}
F.~S. Bemfica, M.~M. Disconzi, and J.~Noronha, ``{Causality and existence of solutions of relativistic viscous fluid dynamics with gravity},'' \href{http://dx.doi.org/10.1103/PhysRevD.98.104064}{{\em Phys. Rev. D} {\bfseries 98} no.~10, (2018) 104064}, \href{http://arxiv.org/abs/1708.06255}{{\ttfamily arXiv:1708.06255 [gr-qc]}}.

\bibitem{Kovtun:FirstorderRelativisticHydrodynamics}
P.~Kovtun, ``{First-order relativistic hydrodynamics is stable},'' \href{http://dx.doi.org/10.1007/JHEP10(2019)034}{{\em JHEP} {\bfseries 10} (2019) 034}, \href{http://arxiv.org/abs/1907.08191}{{\ttfamily arXiv:1907.08191 [hep-th]}}.

\bibitem{Hoult:StableCausalRelativistic}
R.~E. Hoult and P.~Kovtun, ``{Stable and causal relativistic Navier-Stokes equations},'' \href{http://dx.doi.org/10.1007/JHEP06(2020)067}{{\em JHEP} {\bfseries 06} (2020) 067}, \href{http://arxiv.org/abs/2004.04102}{{\ttfamily arXiv:2004.04102 [hep-th]}}.

\bibitem{Abboud:CausalStableFirstorder}
N.~Abboud, E.~Speranza, and J.~Noronha, ``{Causal and stable first-order chiral hydrodynamics},'' \href{http://dx.doi.org/10.1103/PhysRevD.109.094007}{{\em Phys. Rev. D} {\bfseries 109} no.~9, (2024) 094007}, \href{http://arxiv.org/abs/2308.02928}{{\ttfamily arXiv:2308.02928 [hep-th]}}.

\bibitem{Biswas:2022hiv}
R.~Biswas, S.~Mitra, and V.~Roy, ``{An expedition to the islands of stability in the first-order causal hydrodynamics},'' \href{http://dx.doi.org/10.1016/j.physletb.2023.137725}{{\em Phys. Lett. B} {\bfseries 838} (2023) 137725}, \href{http://arxiv.org/abs/2211.11358}{{\ttfamily arXiv:2211.11358 [nucl-th]}}.

\bibitem{Bhattacharyya:2023srn}
S.~Bhattacharyya, S.~Mitra, and S.~Roy, ``{Frame transformation and stable-causal hydrodynamic theory},'' \href{http://arxiv.org/abs/2312.16407}{{\ttfamily arXiv:2312.16407 [nucl-th]}}.

\bibitem{Bhattacharyya:2024tfj}
S.~Bhattacharyya, S.~Mitra, and S.~Roy, ``{Causality and stability in relativistic hydrodynamic theory - a choice to be endured},'' \href{http://dx.doi.org/10.1016/j.physletb.2024.138918}{{\em Phys. Lett. B} {\bfseries 856} (2024) 138918}, \href{http://arxiv.org/abs/2407.18997}{{\ttfamily arXiv:2407.18997 [nucl-th]}}.

\bibitem{Bhattacharyya:2024ohn}
S.~Bhattacharyya, S.~Mitra, S.~Roy, and R.~Singh, ``{Field redefinition and its impact in relativistic hydrodynamics},'' \href{http://arxiv.org/abs/2409.15387}{{\ttfamily arXiv:2409.15387 [nucl-th]}}.

\bibitem{Kovtun:LecturesHydrodynamicFluctuations}
P.~Kovtun, ``{Lectures on hydrodynamic fluctuations in relativistic theories},'' \href{http://dx.doi.org/10.1088/1751-8113/45/47/473001}{{\em J. Phys. A} {\bfseries 45} (2012) 473001}, \href{http://arxiv.org/abs/1205.5040}{{\ttfamily arXiv:1205.5040 [hep-th]}}.

\bibitem{Martin:StatisticalDynamicsClassical}
P.~C. Martin, E.~D. Siggia, and H.~A. Rose, ``Statistical {{Dynamics}} of {{Classical Systems}},''.

\bibitem{Landau:HydrodynamicFluctuations}
L.~D. Landau and E.~M. Lifshitz, ``Hydrodynamic {{Fluctuations}},''.

\bibitem{Hohenberg:TheoryDynamicCritical}
P.~C. Hohenberg and B.~I. Halperin, ``{Theory of Dynamic Critical Phenomena},'' \href{http://dx.doi.org/10.1103/RevModPhys.49.435}{{\em Rev. Mod. Phys.} {\bfseries 49} (1977) 435--479}.

\bibitem{Harder:ThermalFluctuationsGenerating}
M.~Harder, P.~Kovtun, and A.~Ritz, ``{On thermal fluctuations and the generating functional in relativistic hydrodynamics},'' \href{http://dx.doi.org/10.1007/JHEP07(2015)025}{{\em JHEP} {\bfseries 07} (2015) 025}, \href{http://arxiv.org/abs/1502.03076}{{\ttfamily arXiv:1502.03076 [hep-th]}}.

\bibitem{Haehl:TwoRoadsHydrodynamic}
F.~M. Haehl, R.~Loganayagam, and M.~Rangamani, ``{Two roads to hydrodynamic effective actions: a comparison},'' \href{http://arxiv.org/abs/1701.07896}{{\ttfamily arXiv:1701.07896 [hep-th]}}.

\bibitem{Chou:EquilibriumNonequilibriumFormalisms}
K.-c. Chou, Z.-b. Su, B.-l. Hao, and L.~Yu, ``{Equilibrium and Nonequilibrium Formalisms Made Unified},'' \href{http://dx.doi.org/10.1016/0370-1573(85)90136-X}{{\em Phys. Rept.} {\bfseries 118} (1985) 1--131}.

\bibitem{DeSchepper:NonexistenceLinearDiffusion}
I.~M. De~Schepper, H.~Van~Beyeren, and M.~H. Ernst, ``The nonexistence of the linear diffusion equation beyond {{Fick}}'s law,'' \href{http://dx.doi.org/10.1016/0031-8914(74)90290-0}{{\em Physica} {\bfseries 75} no.~1, (1974) 1--36}.

\bibitem{Kovtun:StickinessSoundAbsolute}
P.~Kovtun, G.~D. Moore, and P.~Romatschke, ``{The stickiness of sound: An absolute lower limit on viscosity and the breakdown of second order relativistic hydrodynamics},'' \href{http://dx.doi.org/10.1103/PhysRevD.84.025006}{{\em Phys. Rev. D} {\bfseries 84} (2011) 025006}, \href{http://arxiv.org/abs/1104.1586}{{\ttfamily arXiv:1104.1586 [hep-ph]}}.

\bibitem{Forster:LargedistanceLongtimeProperties}
D.~Forster, D.~R. Nelson, and M.~J. Stephen, ``{Large-distance and long-time properties of a randomly stirred fluid},'' \href{http://dx.doi.org/10.1103/PhysRevA.16.732}{{\em Phys. Rev. A} {\bfseries 16} (1977) 732--749}.

\bibitem{Pomeau:TimeDependentCorrelation}
Y.~Pomeau and P.~Resibois, ``{Time Dependent Correlation Functions and Mode-Mode Coupling Theories},'' {\em Physics Reports} {\bfseries 19} (10, 1974) 63--139.

\bibitem{Santos:DivergenceChapmanEnskogExpansion}
G.~S. Denicol and J.~Noronha, ``{Divergence of the Chapman-Enskog expansion in relativistic kinetic theory},'' \href{http://arxiv.org/abs/1608.07869}{{\ttfamily arXiv:1608.07869 [nucl-th]}}.

\bibitem{Heller:HydrodynamizationKineticTheory}
M.~P. Heller, A.~Kurkela, M.~Spali\'nski, and V.~Svensson, ``{Hydrodynamization in kinetic theory: Transient modes and the gradient expansion},'' \href{http://dx.doi.org/10.1103/PhysRevD.97.091503}{{\em Phys. Rev. D} {\bfseries 97} no.~9, (2018) 091503}, \href{http://arxiv.org/abs/1609.04803}{{\ttfamily arXiv:1609.04803 [nucl-th]}}.

\bibitem{Romatschke:FluidDynamicsFar}
P.~Romatschke, ``{Relativistic Fluid Dynamics Far From Local Equilibrium},'' \href{http://dx.doi.org/10.1103/PhysRevLett.120.012301}{{\em Phys. Rev. Lett.} {\bfseries 120} no.~1, (2018) 012301}, \href{http://arxiv.org/abs/1704.08699}{{\ttfamily arXiv:1704.08699 [hep-th]}}.

\bibitem{Bu:LinearizedFluidGravity}
Y.~Bu and M.~Lublinsky, ``{Linearized fluid/gravity correspondence: from shear viscosity to all order hydrodynamics},'' \href{http://dx.doi.org/10.1007/JHEP11(2014)064}{{\em JHEP} {\bfseries 11} (2014) 064}, \href{http://arxiv.org/abs/1409.3095}{{\ttfamily arXiv:1409.3095 [hep-th]}}.

\bibitem{Huang:StatisticalMechanics}
K.~Huang, {\em Statistical {{Mechanics}}}.
\newblock {Wiley}, 1988.

\bibitem{Denicol:MicroscopicFoundationsRelativistic}
G.~S. Denicol and D.~H. Rischke, {\em Microscopic {{Foundations}} of {{Relativistic Fluid Dynamics}}}, vol.~990 of {\em Lecture {{Notes}} in {{Physics}}}.
\newblock {Springer}, 2022.

\bibitem{Denicol:DerivationTransientRelativistic}
G.~S. Denicol, H.~Niemi, E.~Molnar, and D.~H. Rischke, ``{Derivation of transient relativistic fluid dynamics from the Boltzmann equation},'' \href{http://dx.doi.org/10.1103/PhysRevD.85.114047}{{\em Phys. Rev. D} {\bfseries 85} (2012) 114047}, \href{http://arxiv.org/abs/1202.4551}{{\ttfamily arXiv:1202.4551 [nucl-th]}}. [Erratum: Phys.Rev.D 91, 039902 (2015)].

\bibitem{Rocha:NovelRelaxationTime}
G.~S. Rocha, G.~S. Denicol, and J.~Noronha, ``{Novel Relaxation Time Approximation to the Relativistic Boltzmann Equation},'' \href{http://dx.doi.org/10.1103/PhysRevLett.127.042301}{{\em Phys. Rev. Lett.} {\bfseries 127} no.~4, (2021) 042301}, \href{http://arxiv.org/abs/2103.07489}{{\ttfamily arXiv:2103.07489 [nucl-th]}}.

\bibitem{Hoult:CausalFirstorderHydrodynamics}
R.~E. Hoult and P.~Kovtun, ``{Causal first-order hydrodynamics from kinetic theory and holography},'' \href{http://dx.doi.org/10.1103/PhysRevD.106.066023}{{\em Phys. Rev. D} {\bfseries 106} no.~6, (2022) 066023}, \href{http://arxiv.org/abs/2112.14042}{{\ttfamily arXiv:2112.14042 [hep-th]}}.

\bibitem{Basar:2024qxd}
G.~Ba\c{s}ar, J.~Bhambure, R.~Singh, and D.~Teaney, ``{The stochastic relativistic advection diffusion equation from the Metropolis algorithm},'' \href{http://arxiv.org/abs/2403.04185}{{\ttfamily arXiv:2403.04185 [nucl-th]}}.

\bibitem{Lazo:ActionPrincipleActiondependent}
M.~J. Lazo, J.~Paiva, J.~a. T.~S. Amaral, and G.~a. S.~F. Frederico, ``An {{Action Principle}} for {{Action-dependent Lagrangians}}: toward an {{Action Principle}} to non-conservative systems,'' \href{http://dx.doi.org/10.1063/1.5019936}{{\em Journal of Mathematical Physics} {\bfseries 59} no.~3, (2018) 032902}, \href{http://arxiv.org/abs/1803.08308}{{\ttfamily 1803.08308}}.

\bibitem{Dubovsky:EffectiveFieldTheory}
S.~Dubovsky, L.~Hui, A.~Nicolis, and D.~T. Son, ``{Effective field theory for hydrodynamics: thermodynamics, and the derivative expansion},'' \href{http://dx.doi.org/10.1103/PhysRevD.85.085029}{{\em Phys. Rev. D} {\bfseries 85} (2012) 085029}, \href{http://arxiv.org/abs/1107.0731}{{\ttfamily arXiv:1107.0731 [hep-th]}}.

\bibitem{Dubovsky:EffectiveFieldTheorya}
S.~Dubovsky, L.~Hui, and A.~Nicolis, ``{Effective field theory for hydrodynamics: Wess-Zumino term and anomalies in two spacetime dimensions},'' \href{http://dx.doi.org/10.1103/PhysRevD.89.045016}{{\em Phys. Rev. D} {\bfseries 89} no.~4, (2014) 045016}, \href{http://arxiv.org/abs/1107.0732}{{\ttfamily arXiv:1107.0732 [hep-th]}}.

\bibitem{Jensen:HydrodynamicsEntropyCurrent}
K.~Jensen, M.~Kaminski, P.~Kovtun, R.~Meyer, A.~Ritz, and A.~Yarom, ``{Towards hydrodynamics without an entropy current},'' \href{http://dx.doi.org/10.1103/PhysRevLett.109.101601}{{\em Phys. Rev. Lett.} {\bfseries 109} (2012) 101601}, \href{http://arxiv.org/abs/1203.3556}{{\ttfamily arXiv:1203.3556 [hep-th]}}.

\bibitem{Banerjee:ConstraintsFluidDynamics}
N.~Banerjee, J.~Bhattacharya, S.~Bhattacharyya, S.~Jain, S.~Minwalla, and T.~Sharma, ``{Constraints on Fluid Dynamics from Equilibrium Partition Functions},'' \href{http://dx.doi.org/10.1007/JHEP09(2012)046}{{\em JHEP} {\bfseries 09} (2012) 046}, \href{http://arxiv.org/abs/1203.3544}{{\ttfamily arXiv:1203.3544 [hep-th]}}.

\bibitem{Jain:SchwingerKeldyshEffectiveField}
A.~Jain and P.~Kovtun, ``{Schwinger-Keldysh effective field theory for stable and causal relativistic hydrodynamics},'' \href{http://dx.doi.org/10.1007/JHEP01(2024)162}{{\em JHEP} {\bfseries 01} (2024) 162}, \href{http://arxiv.org/abs/2309.00511}{{\ttfamily arXiv:2309.00511 [hep-th]}}.

\bibitem{Armas:EffectiveFieldTheory}
J.~Armas and A.~Jain, ``{Effective field theory for hydrodynamics without boosts},'' \href{http://dx.doi.org/10.21468/SciPostPhys.11.3.054}{{\em SciPost Phys.} {\bfseries 11} no.~3, (2021) 054}, \href{http://arxiv.org/abs/2010.15782}{{\ttfamily arXiv:2010.15782 [hep-th]}}.

\bibitem{Jain:EffectiveFieldTheory}
A.~Jain, ``{Effective field theory for non-relativistic hydrodynamics},'' \href{http://dx.doi.org/10.1007/JHEP10(2020)208}{{\em JHEP} {\bfseries 10} (2020) 208}, \href{http://arxiv.org/abs/2008.03994}{{\ttfamily arXiv:2008.03994 [hep-th]}}.

\bibitem{Maldacena:LargeLimitSuperconformal}
J.~M. Maldacena, ``{The Large N limit of superconformal field theories and supergravity},'' \href{http://dx.doi.org/10.4310/ATMP.1998.v2.n2.a1}{{\em Adv. Theor. Math. Phys.} {\bfseries 2} (1998) 231--252}, \href{http://arxiv.org/abs/hep-th/9711200}{{\ttfamily arXiv:hep-th/9711200}}.

\bibitem{Gubser:GaugeTheoryCorrelators}
S.~S. Gubser, I.~R. Klebanov, and A.~M. Polyakov, ``{Gauge theory correlators from noncritical string theory},'' \href{http://dx.doi.org/10.1016/S0370-2693(98)00377-3}{{\em Phys. Lett. B} {\bfseries 428} (1998) 105--114}, \href{http://arxiv.org/abs/hep-th/9802109}{{\ttfamily arXiv:hep-th/9802109}}.

\bibitem{Witten:SitterSpaceHolography}
E.~Witten, ``{Anti-de Sitter space and holography},'' \href{http://dx.doi.org/10.4310/ATMP.1998.v2.n2.a2}{{\em Adv. Theor. Math. Phys.} {\bfseries 2} (1998) 253--291}, \href{http://arxiv.org/abs/hep-th/9802150}{{\ttfamily arXiv:hep-th/9802150}}.

\bibitem{Thorne:BlackHolesMembrane}
K.~S. Thorne, R.~H. Price, and D.~A. MacDonald, {\em Black {{Holes}}: {{The Membrane Paradigm}}}.
\newblock The {{Silliman Memorial Lectures Series}}. {Yale University Press}, 1986.

\bibitem{Iqbal:UniversalityHydrodynamicLimit}
N.~Iqbal and H.~Liu, ``{Universality of the hydrodynamic limit in AdS/CFT and the membrane paradigm},'' \href{http://dx.doi.org/10.1103/PhysRevD.79.025023}{{\em Phys. Rev. D} {\bfseries 79} (2009) 025023}, \href{http://arxiv.org/abs/0809.3808}{{\ttfamily arXiv:0809.3808 [hep-th]}}.

\bibitem{Policastro:AdSCFTCorrespondence}
G.~Policastro, D.~T. Son, and A.~O. Starinets, ``{From AdS / CFT correspondence to hydrodynamics},'' \href{http://dx.doi.org/10.1088/1126-6708/2002/09/043}{{\em JHEP} {\bfseries 09} (2002) 043}, \href{http://arxiv.org/abs/hep-th/0205052}{{\ttfamily arXiv:hep-th/0205052}}.

\bibitem{Policastro:AdSCFTCorrespondencea}
G.~Policastro, D.~T. Son, and A.~O. Starinets, ``{From AdS / CFT correspondence to hydrodynamics. 2. Sound waves},'' \href{http://dx.doi.org/10.1088/1126-6708/2002/12/054}{{\em JHEP} {\bfseries 12} (2002) 054}, \href{http://arxiv.org/abs/hep-th/0210220}{{\ttfamily arXiv:hep-th/0210220}}.

\bibitem{Ramallo:IntroductionAdSCFT}
A.~V. Ramallo, ``{Introduction to the AdS/CFT correspondence},'' \href{http://dx.doi.org/10.1007/978-3-319-12238-0_10}{{\em Springer Proc. Phys.} {\bfseries 161} (2015) 411--474}, \href{http://arxiv.org/abs/1310.4319}{{\ttfamily arXiv:1310.4319 [hep-th]}}.

\bibitem{Natsuume:AdSCFTDuality}
M.~Natsuume, \href{http://dx.doi.org/10.1007/978-4-431-55441-7}{{\em {AdS/CFT Duality User Guide}}}, vol.~903.
\newblock Springer, 2015.
\newblock \href{http://arxiv.org/abs/1409.3575}{{\ttfamily arXiv:1409.3575 [hep-th]}}.

\bibitem{Kovtun:HolographyHydrodynamicsDiffusion}
P.~Kovtun, D.~T. Son, and A.~O. Starinets, ``{Holography and hydrodynamics: Diffusion on stretched horizons},'' \href{http://dx.doi.org/10.1088/1126-6708/2003/10/064}{{\em JHEP} {\bfseries 10} (2003) 064}, \href{http://arxiv.org/abs/hep-th/0309213}{{\ttfamily arXiv:hep-th/0309213}}.

\bibitem{Singh:2024qvg}
R.~Singh, ``{On the non-uniqueness of the energy-momentum and spin currents},'' \href{http://arxiv.org/abs/2406.02127}{{\ttfamily arXiv:2406.02127 [hep-th]}}.

\bibitem{Israel:ThermodynamicsRelativisticSystems}
W.~Israel, ``Thermodynamics of relativistic systems,''.

\bibitem{Armas:CarrollianFluidsSpontaneous}
J.~Armas and E.~Have, ``{Carrollian Fluids and Spontaneous Breaking of Boost Symmetry},'' \href{http://dx.doi.org/10.1103/PhysRevLett.132.161606}{{\em Phys. Rev. Lett.} {\bfseries 132} no.~16, (2024) 161606}, \href{http://arxiv.org/abs/2308.10594}{{\ttfamily arXiv:2308.10594 [hep-th]}}.

\bibitem{Eckart:ThermodynamicsIrreversibleProcesses}
C.~Eckart, ``The {{Thermodynamics}} of {{Irreversible Processes}}. {{III}}. {{Relativistic Theory}} of the {{Simple Fluid}},''.

\bibitem{Landau:StatisticalPhysicsVolume}
L.~D. Landau and E.~M. Lifshitz, {\em Statistical {{Physics}}: {{Volume}} 5}.
\newblock Landau and {{Lifshitz}}: {{Course}} of {{Theoretical Physics}}. {Elsevier}, 1980.

\bibitem{Endlich:DissipationEffectiveField}
S.~Endlich, A.~Nicolis, R.~A. Porto, and J.~Wang, ``{Dissipation in the effective field theory for hydrodynamics: First order effects},'' \href{http://dx.doi.org/10.1103/PhysRevD.88.105001}{{\em Phys. Rev. D} {\bfseries 88} (2013) 105001}, \href{http://arxiv.org/abs/1211.6461}{{\ttfamily arXiv:1211.6461 [hep-th]}}.

\bibitem{Dore:FluctuatingRelativisticDissipative}
T.~Dore, L.~Gavassino, D.~Montenegro, M.~Shokri, and G.~Torrieri, ``{Fluctuating relativistic dissipative hydrodynamics as a gauge theory},'' \href{http://dx.doi.org/10.1016/j.aop.2022.168902}{{\em Annals Phys.} {\bfseries 442} (2022) 168902}, \href{http://arxiv.org/abs/2109.06389}{{\ttfamily arXiv:2109.06389 [hep-th]}}.

\bibitem{Armas:2020mpr}
J.~Armas and A.~Jain, ``{Effective field theory for hydrodynamics without boosts},'' \href{http://dx.doi.org/10.21468/SciPostPhys.11.3.054}{{\em SciPost Phys.} {\bfseries 11} no.~3, (2021) 054}, \href{http://arxiv.org/abs/2010.15782}{{\ttfamily arXiv:2010.15782 [hep-th]}}.

\bibitem{Bhattacharyya:EntropyCurrentPartition}
S.~Bhattacharyya, ``{Entropy Current from Partition Function: One Example},'' \href{http://dx.doi.org/10.1007/JHEP07(2014)139}{{\em JHEP} {\bfseries 07} (2014) 139}, \href{http://arxiv.org/abs/1403.7639}{{\ttfamily arXiv:1403.7639 [hep-th]}}.

\bibitem{Bhattacharyya:EntropyCurrentEquilibrium}
S.~Bhattacharyya, ``{Entropy current and equilibrium partition function in fluid dynamics},'' \href{http://dx.doi.org/10.1007/JHEP08(2014)165}{{\em JHEP} {\bfseries 08} (2014) 165}, \href{http://arxiv.org/abs/1312.0220}{{\ttfamily arXiv:1312.0220 [hep-th]}}.

\bibitem{Kurkela:2017xis}
A.~Kurkela and U.~A. Wiedemann, ``{Analytic structure of nonhydrodynamic modes in kinetic theory},'' \href{http://dx.doi.org/10.1140/epjc/s10052-019-7271-9}{{\em Eur. Phys. J. C} {\bfseries 79} no.~9, (2019) 776}, \href{http://arxiv.org/abs/1712.04376}{{\ttfamily arXiv:1712.04376 [hep-ph]}}.

\bibitem{Ochsenfeld:2023wxz}
S.~Ochsenfeld and S.~Schlichting, ``{Hydrodynamic and non-hydrodynamic excitations in kinetic theory \textemdash{} a numerical analysis in scalar field theory},'' \href{http://dx.doi.org/10.1007/JHEP09(2023)186}{{\em JHEP} {\bfseries 09} (2023) 186}, \href{http://arxiv.org/abs/2308.04491}{{\ttfamily arXiv:2308.04491 [hep-th]}}.

\bibitem{Bajec:2024jez}
M.~Bajec, S.~Grozdanov, and A.~Soloviev, ``{Spectra of correlators in the relaxation time approximation of kinetic theory},'' \href{http://dx.doi.org/10.1007/JHEP08(2024)065}{{\em JHEP} {\bfseries 08} (2024) 065}, \href{http://arxiv.org/abs/2403.17769}{{\ttfamily arXiv:2403.17769 [hep-th]}}.

\bibitem{Brants:2024wrx}
R.~Brants, ``{New insights into the analytic structure of correlation functions via kinetic theory},'' \href{http://dx.doi.org/10.1103/PhysRevD.110.116027}{{\em Phys. Rev. D} {\bfseries 110} no.~11, (2024) 116027}, \href{http://arxiv.org/abs/2409.09022}{{\ttfamily arXiv:2409.09022 [hep-th]}}.

\bibitem{Kadanoff:HydrodynamicEquationsCorrelation}
L.~P. Kadanoff and P.~C. Martin, ``Hydrodynamic equations and correlation functions,''.

\bibitem{Herzog:LecturesHolographicSuperfluidity}
C.~P. Herzog, ``{Lectures on Holographic Superfluidity and Superconductivity},'' \href{http://dx.doi.org/10.1088/1751-8113/42/34/343001}{{\em J. Phys. A} {\bfseries 42} (2009) 343001}, \href{http://arxiv.org/abs/0904.1975}{{\ttfamily arXiv:0904.1975 [hep-th]}}.

\bibitem{Onsager:ReciprocalRelationsIrreversible}
L.~Onsager, ``Reciprocal {{Relations}} in {{Irreversible Processes}}. {{I}}.,'' \href{http://dx.doi.org/10.1103/PhysRev.37.405}{{\em Phys. Rev.} {\bfseries 37} no.~4, }.

\bibitem{Onsager:ReciprocalRelationsIrreversiblea}
L.~Onsager, ``Reciprocal relations in irreversible processes. {{II}}.,'' \href{http://dx.doi.org/10.1103/PhysRev.38.2265}{{\em Phys. Rev.} {\bfseries 38} no.~12, }.

\bibitem{Ammon:ChiralHydrodynamicsStrong}
M.~Ammon, S.~Grieninger, J.~Hernandez, M.~Kaminski, R.~Koirala, J.~Leiber, and J.~Wu, ``{Chiral hydrodynamics in strong external magnetic fields},'' \href{http://dx.doi.org/10.1007/JHEP04(2021)078}{{\em JHEP} {\bfseries 04} (2021) 078}, \href{http://arxiv.org/abs/2012.09183}{{\ttfamily arXiv:2012.09183 [hep-th]}}.

\bibitem{Hartnoll:LecturesHolographicMethods}
S.~A. Hartnoll, ``{Lectures on holographic methods for condensed matter physics},'' \href{http://dx.doi.org/10.1088/0264-9381/26/22/224002}{{\em Class. Quant. Grav.} {\bfseries 26} (2009) 224002}, \href{http://arxiv.org/abs/0903.3246}{{\ttfamily arXiv:0903.3246 [hep-th]}}.

\bibitem{Amoretti:2020mkp}
A.~Amoretti, D.~K. Brattan, N.~Magnoli, and M.~Scanavino, ``{Magneto-thermal transport implies an incoherent Hall conductivity},'' \href{http://dx.doi.org/10.1007/JHEP08(2020)097}{{\em JHEP} {\bfseries 08} (2020) 097}, \href{http://arxiv.org/abs/2005.09662}{{\ttfamily arXiv:2005.09662 [hep-th]}}.

\bibitem{Amoretti:2023hpb}
A.~Amoretti, D.~K. Brattan, L.~Martinoia, I.~Matthaiakakis, and J.~Rongen, ``{Relaxation terms for anomalous hydrodynamic transport in Weyl semimetals from kinetic theory},'' \href{http://dx.doi.org/10.1007/JHEP02(2024)071}{{\em JHEP} {\bfseries 02} (2024) 071}, \href{http://arxiv.org/abs/2309.05692}{{\ttfamily arXiv:2309.05692 [hep-th]}}.

\bibitem{Amoretti:2023vhe}
A.~Amoretti, D.~K. Brattan, L.~Martinoia, and I.~Matthaiakakis, ``{Restoring time-reversal covariance in relaxed hydrodynamics},'' \href{http://dx.doi.org/10.1103/PhysRevD.108.056003}{{\em Phys. Rev. D} {\bfseries 108} no.~5, (2023) 056003}, \href{http://arxiv.org/abs/2304.01248}{{\ttfamily arXiv:2304.01248 [hep-th]}}.

\bibitem{Singh:2021man}
R.~Singh, M.~Shokri, and R.~Ryblewski, ``{Spin polarization dynamics in the Bjorken-expanding resistive MHD background},'' \href{http://dx.doi.org/10.1103/PhysRevD.103.094034}{{\em Phys. Rev. D} {\bfseries 103} no.~9, (2021) 094034}, \href{http://arxiv.org/abs/2103.02592}{{\ttfamily arXiv:2103.02592 [hep-ph]}}.

\bibitem{Singh:2022ltu}
R.~Singh, M.~Shokri, and S.~M. A.~T. Mehr, ``{Relativistic hydrodynamics with spin in the presence of electromagnetic fields},'' \href{http://dx.doi.org/10.1016/j.nuclphysa.2023.122656}{{\em Nucl. Phys. A} {\bfseries 1035} (2023) 122656}, \href{http://arxiv.org/abs/2202.11504}{{\ttfamily arXiv:2202.11504 [hep-ph]}}.

\end{thebibliography}\endgroup
\bibliographystyle{utphys}
\end{document}